\documentclass[journal]{new-aiaa}
\usepackage[utf8]{inputenc}

\usepackage{graphicx}
\usepackage{amsmath}
\usepackage[version=4]{mhchem}
\usepackage{siunitx}
\usepackage{longtable,tabularx}
\usepackage{indentfirst}
\usepackage{hyperref}
\usepackage{mathtools}
\usepackage{bm}
\usepackage{xspace}
\usepackage{wrapfig}
\usepackage{multicol}
\usepackage{xurl}
\usepackage[flushleft]{threeparttable}
\usepackage{booktabs}
\usepackage{gensymb}
\usepackage{mdframed}
\usepackage{subcaption}
\usepackage{xcolor, pifont}
\usepackage{lscape}
\usepackage{tcolorbox}
\usepackage[ruled, vlined, linesnumbered]{algorithm2e}

\let\svthefootnote\thefootnote
\newcommand\freefootnote[1]{
  \let\thefootnote\relax
  \footnotetext{#1}
  \let\thefootnote\svthefootnote
}

\newcommand{\REOSSP}{\hyperlink{REOSSP}{\textsf{REOSSP}}\xspace}

\definecolor{myblue}{rgb}{0, 0.23, 0.64}
\definecolor{WVUblue}{rgb}{0, 0.16, 0.33}

\hypersetup{
    colorlinks=true,
    linkcolor=myblue,
    filecolor=magenta,
    citecolor=myblue,
    urlcolor=myblue,
}

\title{Automating the Wildfire Detection and Scheduling Pipeline with Maneuverable Earth Observation Satellites}

\author{Brycen D. Pearl\footnote{Ph.D. Candidate, Department of Mechanical, Materials and Aerospace Engineering, Student Member AIAA.}, Joshua G. Warner\footnote{Undergraduate Student, Department of Mechanical, Materials and Aerospace Engineering.}, and Hang Woon Lee\footnote{Assistant Professor, Department of Mechanical, Materials and Aerospace Engineering; hangwoon.lee@mail.wvu.edu. Member AIAA (Corresponding Author).}}
\affil{West Virginia University, Morgantown, WV, 26506}

\begin{document}

\newpage

\freefootnote{This paper is a substantially revised version of the paper AAS 25-687, presented at the 2025 AAS/AIAA Astrodynamics Specialist Conference, Boston, MA, August 10-14, 2025. It offers substantial revisions to the components of the developed framework, a new instance of experimentation with refined results, and a better description of the materials.}

\maketitle

\begin{abstract} 
    Wildfires are becoming increasingly frequent, with potentially devastating consequences, including loss of life, infrastructure destruction, and severe environmental damage. Low-Earth-orbit satellites equipped with onboard sensors can capture critical information related to active wildfires and enable near-real-time detection through machine learning algorithms applied to the acquired data. We propose a framework that automates the complete wildfire detection and satellite scheduling pipeline, entitled the WildFire-applicable Intelligent and Responsive Ensemble for Detection and Scheduling (WildFIRE-DS). This paper develops an algorithm to realize the vision of the WildFIRE-DS as a proof of concept, integrating three key components: wildfire detection in satellite imagery, statistical updating that incorporates data from repeated flyovers, and multisatellite scheduling optimization. The algorithm enables wildfire detection using convolutional neural networks with sensor fusion techniques, incorporates subsequent flyover information via Bayesian statistics, and schedules a constellation of satellites using the state-of-the-art Reconfigurable Earth Observation Satellite Scheduling Problem. Simulated experiments conducted using real-world wildfire locations and the orbits of operational Earth observation satellites demonstrate that this autonomous detection and scheduling approach effectively enhances wildfire monitoring capabilities.
\end{abstract}

\section*{Nomenclature}
{
\renewcommand\arraystretch{1.0}
\noindent\begin{longtable*}{@{}l @{\quad=\quad} l@{}}
$B_{\text{various}}$ & battery-related values \\
$b$                  & indicator variable of the current battery storage level \\
$C$                  & data downlink weight in Schedule Module objective functions \\
$c$                  & cost of orbital maneuver \\
$c_{\max}$           & budget for total orbital maneuver costs \\
$D_{\text{various}}$ & data-related values \\
$d$                  & indicator variable of the current data storage level \\
$F$                  & F-score of convolutional neural network model \\
$\mathcal{G}$        & set of ground stations, index $g$, cardinality $G$ \\
$H$                  & binary visibility condition of the sun \\
$h$                  & decision variable to control solar charging \\
$\mathcal{J}$        & set of orbital slot options, indices $i,j$, cardinality $J$ \\
$\mathcal{K}$        & set of satellites, index $k$, cardinality $K$ \\
$O$                  & auxiliary target visibility weighting in objective functions \\
$\mathcal{P}$        & set of priority targets, index $p$, cardinality $P$ \\
$\mathcal{P}^\prime$ & set of auxiliary targets, index $p$, cardinality $P^\prime$ \\
$P(X|Y)$             & the probability that $X$ occurs, given $Y$ occurs \\
$P_r$                & precision of convolutional neural network model \\
$q$                  & decision variable to control data downlink to ground stations \\
$R$                  & recall of convolutional neural network model \\
$\mathcal{S}$        & set of reconfiguration stages, index $s$, cardinality $S+1$ \\
$\mathcal{T}$        & set of time steps, index $t$, cardinality $T$ \\
$T_r$                & finite schedule duration \\
$U$                  & binary visibility condition of auxiliary targets \\
$V$                  & binary visibility condition of priority targets \\
$v_z$                & vertical component of satellite velocity \\
$W$                  & binary visibility condition of ground stations \\
$x$                  & decision variable to control constellation reconfiguration \\
$y$                  & decision variable to control priority target observation \\
$z$                  & objective function value \\
$\alpha$             & decision variable to control obtained auxiliary visibility \\
$\Delta t$           & time step size between discrete time steps \\
$\delta x_{\text{Pixels}}, \delta y_{\text{Pixels}}$ & distance from image center to detected wildfire in pixels \\
$\delta\text{lat}, \delta\text{lon}$ & distance from satellite latitude/longitude to detected wildfire in degrees latitude/longitude \\
$\theta$             & angle from east direction to direction of satellite travel, satellite inclination \\
$\phi$               & angle from image right-facing direction to detected wildfire \\
$\psi$               & angle from east direction to direction of detection wildfire \\
\end{longtable*}
}

\section{Introduction}

Wildfires are widespread, frequent natural processes that span a variety of spatial and temporal scales; particularly extensive or rapidly spreading wildfires are classified as natural disasters, thus posing a significant threat to human life and developed infrastructure. The impacts of wildfires vary widely, including loss of life, burned area \cite{Jain2020ML}, the spread of smoke in populated areas \cite{Bayham2022Annual,Jain2020ML}, and the destruction of property, resulting in negative effects on the economy \cite{Bayham2022Annual}. Additionally, wildfire disasters have increased in frequency and size in recent years \cite{Oliveira2021Wildfire}, including (but not limited to) the Australian ``Black Summer'' wildfires of 2019 through 2020 \cite{Davey2020Australia}, the California wildfires of 2020 \cite{Keeley2021California}, the Canadian wildfires of 2023 \cite{Pelletier2024Canada}, and the 2025 South Korea wildfires \cite{Sung2025SouthKorea}, all of which are the largest in history for their respective regions. More recently, the Palisades Fire and the Eaton Fire, both of which occurred in California in January of 2025, are the largest wildfires to occur in the history of Los Angeles. The Palisades Fire burned a total of \num{23448} acres, killed $12$ people, destroyed \num{6837} structures, and caused \$$25$ billion in damages \cite{Flores2025Training,CalFIrePalisades}, while the Eaton Fire similarly burned a total of \num{14021} acres, killed $19$ people, destroyed \num{9414} structures, and caused \$$27.5$ billion in damages \cite{Han2025Spatial,CalFIreEaton} during the same time frame. Moreover, the progression of wildfires is a highly complex process; humans are responsible for over \SI{90}{\%} of wildfire ignitions, and naturally occurring lightning strikes account for a majority of the remaining \SI{10}{\%} \cite{Jain2020ML}. As such, the threat that wildfires pose lies in their unpredictable nature; therefore, additional information, such as temperature and wind direction data, would allow first responders and relief agencies to make adequate preparations or respond appropriately, as well as allow crucial decisions to be made earlier \cite{Sung2025SouthKorea}. Thus, wildfire detection and monitoring techniques must be advanced to obtain critical wildfire information more rapidly.

Various observation systems are employed to identify and gather data on wildfires, including those conducted on the ground, in the air, and from space. Concerning ground-based systems, sensor nodes measuring pressure, temperature, humidity, and/or wind, such as Dryad's Silvanet Mesh Gateway \cite{Mohapatra2022Silvanet}, have been leveraged to detect wildfires \cite{Udaya2022ML}. Similarly, manned and unmanned aerial vehicles (UAVs) can reach otherwise inaccessible vantage points above a wildfire and have been used to gather position and temperature data to interpret wildfire positions \cite{Zhongjie2019Kalman} and detect wildfires through visible and infrared measurements \cite{Tomasz2021Multimodal}. Separately, the use of low-Earth-orbit (LEO) satellites equipped with various sensors is a prominent technique for observing and detecting natural disasters \cite{boustan2020}, including wildfires. While UAVs can obtain inaccessible vantage points, satellite systems improve upon UAV effectiveness through expanded coverage and frequent revisit rates while also maintaining or improving upon resolution with ground sample distances on the order of meters per pixel \cite{Sandau2007GSD}. Additionally, constellations of satellites can cooperatively revisit natural disaster locations to detect spatial and temporal changes \cite{Voigt2007Satellite}, without the need for subsequent deployment or management by on-site personnel. These aspects of satellite operation further emphasize their usefulness in observing wildfires.

A variety of instruments have been employed on LEO satellites for wildfire detection. For instance, the Advanced Very High Resolution Radiometer (AVHRR), a multispectral scanner with a medium spatial resolution of \SI{1.1}{km} \cite{Huh1991Limitations} designed to gather data on vegetation conditions and surface temperature onboard the NOAA-18 and NOAA-19 satellites \cite{Kalluri2021AVHRR}, is used to detect and monitor the progression of active or possible wildfires \cite{Domenikiotis2003AVHRRUse}. Moreover, the Moderate Resolution Imaging Spectroradiometer (MODIS), utilizing $36$ spectral bands \cite{MODISDescription} onboard NASA's Terra and Aqua satellites, is used to obtain imagery of wildfires \cite{Veraverbeke2014MODISFire} due to its wide spectral range. Recently, the Visible Infrared Imaging Radiometer Suite (VIIRS), providing observations across $22$ spectral bands \cite{Cao2025VIIRS} onboard NOAA-20 and NOAA-21, has been used to algorithmically detect active wildfires \cite{Zhang2023WildfireDetection}, further providing data to the NASA Land, Atmosphere Near-Real-time Capability for Earth Observation \cite{NASALANCE} Fire Information for Resource Management System (FIRMS) \cite{NASAFIRMS}.

Recent applications of satellite imagery for detecting active wildfires have favored the use of analytical algorithms, while others employ more flexible machine learning methods to overcome the limitations of analytical implementations. Some prominent uses of analytical algorithms include the use of synthetic aperture radar imagery to determine the fuel load of given forested regions \cite{Saatchi2007Estimation} and the use of edge error quantification to detect and refine the boundaries of burn scars in MODIS imagery \cite{Humber2020Assessing}. Analytical algorithms have respective strengths and are computationally efficient, although they may lack adaptability across varying environmental conditions, sensor types, or present fire characteristics, while additionally struggling to handle the large volumes of data generated by satellite imagery, which often includes complex and high-dimensional information \cite{Jarrallah2022}. To overcome such potential limitations, machine learning methods such as convolutional neural networks (CNNs) have been used as a powerful alternative to analytical algorithms due to the inherent capability to learn complex spatial and spectral patterns. CNNs have been successfully applied to a wide range of remote sensing tasks, including land-cover classification, change detection, and fire pixel identification \cite{Jarrallah2022}.

Once a wildfire is detected within satellite imagery, satellites must be scheduled to optimally collect further data regarding the wildfire, thus requiring the consideration of the Earth observation satellite scheduling problem (EOSSP). The EOSSP is an optimization problem in which each satellite task is treated as a decision variable to maximize rewards related to observing given targets. The EOSSP is commonly expanded using concepts of satellite operation, such as satellite agility to incorporate slewing in the agile EOSSP (AEOSSP), to improve the overall effectiveness of a resultant optimal schedule \cite{Wang2023DRL,Mantovani2026Improving,Khatir2025Enhancing}. An additional concept of satellite operation that has recently been expanded upon is constellation reconfigurability, the capability of satellites within a constellation to perform orbital maneuvers to form a more optimal formation \cite{deweck2008optimal}, to further improve the effectiveness of Earth observation satellites in LEO \cite{Lee2023Regional,Lee2022Maximizing,Pearl2024Benchmarking}. Furthermore, state-of-the-art developments have extended the EOSSP through the use of constellation reconfigurability to create the \textit{Reconfigurable Earth Observation Satellite Scheduling Problem} (\REOSSP) \cite{Pearl2025Developing,Pearl2025REOSSP}, leveraging mixed-integer linear programming (MILP) to increase the amount of observation data obtained in a scheduled mission. 

While active wildfire detection has seen many improvements in recent literature, as presented in Sec.~\ref{sec:Lit_Review}, and satellite scheduling methods have been improved via the implementation of constellation reconfigurability \cite{Pearl2025Developing,Pearl2025REOSSP}, such improved methods have yet to be used in combination with one another. Prior efforts to combine detection and scheduling within a cohesive pipeline, as discussed in Sec.~\ref{subsec:Lit_Pipeline}, rely on outdated algorithms \cite{Davies2006ASE} or autonomously retask a constellation with a significant lead time \cite{Kangaslahti2024Sensorweb}. Overall, given the limitations of these initial efforts, alongside the possibility that dynamic targets such as wildfires may have changed drastically, it is likely that uplinked schedules are outdated. With the devastating impact of wildfires and recent developments in wildfire detection, as discussed in Sec.~\ref{subsec:Lit_CNN}, and satellite scheduling, the motivation to automate the sequential and cyclical process of improved detection and subsequent scheduling aspects of mission planning presents itself. Additional literature covered in the latter portion of Sec.~\ref{subsec:Lit_Pipeline} addresses important aspects of the detection and scheduling pipeline while remaining fragmented and not combining operations to cover the pipeline in its entirety with these recent developments. As such, a framework that incorporates these newly improved aspects of the pipeline for highly dynamic events observed by satellites in LEO with effective spatial and temporal resolution allows for a more effective and rapid response. 

\begin{figure}[!ht]
    \centering
    \includegraphics[width = \textwidth]{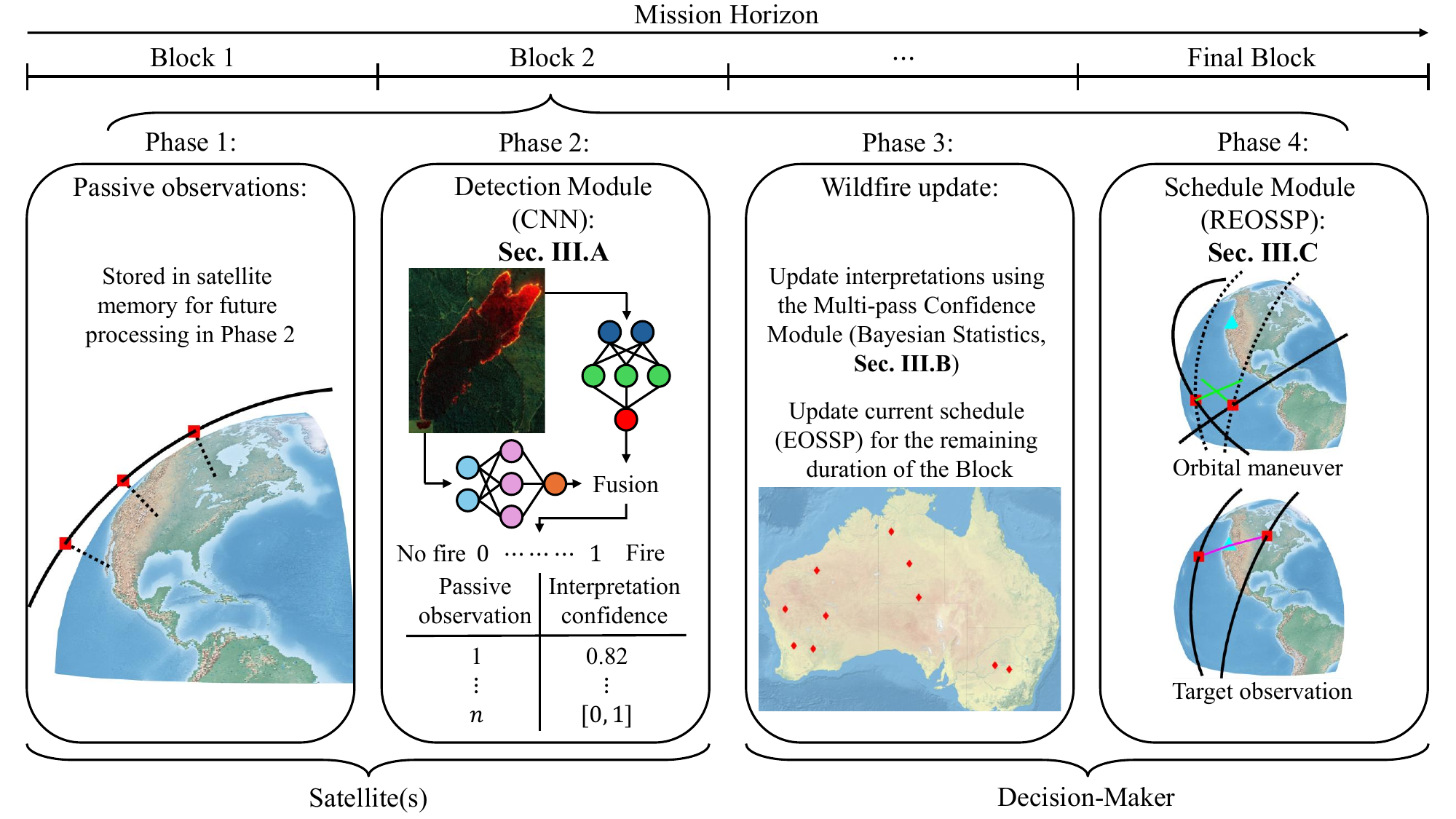}
    \caption{A demonstration of the capabilities within the WildFIRE-DS.}
    \label{fig:Illustrative_Figure}
\end{figure}

A cohesive framework to automate the detection of active wildfires and the scheduling of a constellation of maneuverable satellites is proposed, given recent breakthroughs in wildfire detection and satellite scheduling, as well as the necessity of combining such areas. The proposed framework comprises three key components: wildfire detection, target verification, and satellite scheduling, all of which must be cohesively linked for autonomous operation. This paper develops an algorithm, the WildFire-applicable Intelligent and Responsive Ensemble for Detecting and Scheduling (WildFIRE-DS), to realize the proposed framework. The WildFIRE-DS algorithm bridges the gap between autonomous wildfire detection and satellite scheduling by employing two recently developed methods: CNNs with sensor fusion techniques for detecting wildfires in satellite imagery and the state-of-the-art \REOSSP for scheduling a reconfigurable constellation. Additionally, WildFIRE-DS is depicted in Fig.~\ref{fig:Illustrative_Figure} as a high-level overview. This paper expands upon initial research conducted in Ref.~\cite{Pearl2025Autonomous}, improving the effectiveness of wildfire detection and the responsiveness of satellite scheduling. The results of simulated experimentation in this paper demonstrate that the WildFIRE-DS is effective at conducting autonomous operations for real-world wildfire events, successfully automating the detection and scheduling pipeline for wildfire natural disasters, thus realizing the proposed framework.

The rest of the paper is organized as follows. First, Sec.~\ref{sec:Lit_Review} reviews related literature, explaining concepts used throughout the developed WildFIRE-DS algorithm, including CNN-based object detection and improvements. Then, Sec.~\ref{sec:WildFIRE-DS} details the WildFIRE-DS for the autonomous detection and subsequent scheduling of wildfires, where each subsection depicts a module involved. Next, Sec.~\ref{sec:Experiment} presents a simulated experimental application of the developed algorithm relative to real-world wildfire data and Earth observation satellite orbits. Finally, Sec.~\ref{sec:Conclusion} reviews the results of the experiment, provides concluding remarks, and discusses potentially fruitful future research endeavors.

\section{Literature Review} \label{sec:Lit_Review}

This section details information utilized throughout the rest of the paper and the developed WildFIRE-DS algorithm, providing thorough descriptions of methodology and examples or justification from related literature. 

\subsection{CNN Application to Object Detection in Satellite Imagery} \label{subsec:Lit_CNN}

CNNs are machine learning models designed for feature extraction and pattern recognition and are therefore particularly effective for image classification and segmentation \cite{Pereira2021}. CNNs consist of multiple layers that progressively transform a raw input into hierarchical representations, enabling robust detection of spatial features such as edges, textures, and complex patterns \cite{Goodfellow2016}. There is a multitude of related literature applying CNN architectures to wildfire detection and analysis. For instance, Ref.~\cite{Bo2022BASNet} applies salient object detection to burned-area segmentation to improve UAV performance when processing high-resolution images, outperforming state-of-the-art salient object detection methods both quantitatively and qualitatively. Separately, Ref.~\cite{Lv2022Spatial} develops spatial-spectral attention mechanisms and multiscale dilation convolution modules for land-cover change detection in satellite imagery to determine changes upon subsequent flyovers. Additionally, Ref.~\cite{Li2001Autonmatic} utilizes multi-threshold techniques within a novel neural network architecture to classify satellite imagery obtained from the AVHRR system into smoke, cloud, and clear classifications, thus determining wildfire presence. 

In addition to the basic CNN architecture, algorithms such as the You Only Look Once (YOLO) architecture enhance object detection by introducing a grid-like structure that simultaneously predicts object classification certainty and bounding boxes, which enclose detected objects \cite{Ali2024TheYOLO}. This simplification in design enables YOLO to be used for accurate and timely predictions on large quantities of data in near real time. Several related works of literature further expand upon YOLO for object detection within satellite imagery. Similar to Ref.~\cite{Lv2022Spatial}, Ref.~\cite{Peng2023AMFLW-YOLO} employs coordinate attention to capture direction and location information across input channels simultaneously to improve accuracy when applied to general targets in satellite imagery. Additionally, Ref.~\cite{Zhang2025MobileOne} improves upon a YOLO-based object detection algorithm to efficiently detect fire occurrence by reducing certain parameters, implementing a new activation function leveraging funnel activation, and improving the loss function through the use of an efficient intersection-over-union loss function. Separately, Ref.~\cite{Li2021Few} modifies the YOLO architecture using a metafeature extractor to reweight aspects of an input satellite image to provide effective object detection using only a few annotated samples when training the CNN model. Finally, Ref.~\cite{Zhang2024FFCA-YOLO} implements a three-module architecture aimed at improving local area awareness, multiscale feature fusion, and global association across input channels to enhance the weak feature representation of small objects while simultaneously suppressing background objects. 

\subsection{Effective Spectral Wavelengths for Wildfire Detection} \label{subsec:Lit_SWIR}

Within the use of multispectral imagery for wildfire detection, short-wave infrared (SWIR) wavelength bands are heavily favored in both algorithmic and CNN-based applications for two key reasons. First, the peak emission strengths of flames and smoldering fire temperatures are at wavelengths less than \SI{3.00}{\upmu\meter} \cite{Allison2016}, leading pure IR and visible spectra to be saturated or incapable of measuring bright IR emissions \cite{Riggan2008}. Specifically, the wavelengths at which emissions are not strongly absorbed by water vapor include \SI{1.60}{\upmu\meter} and \SI{2.00}{\upmu\meter} \cite{Riggan2008}. Furthermore, Ref.~\cite{Riggan2008} provides an example in which fires in the Cerrado region of central Brazil obtain a maximum emission radiance in the SWIR range at \SI{2.20}{\upmu\meter}, while Ref.~\cite{Riggan2009} provides an additional example wherein a maximum emission radiance of \SI{1.63}{\upmu\meter} is obtained. Second, SWIR is more practical than visible spectra both at night, due to direct measurement of flame temperature emission, and during the day, due to better haze and smoke penetration \cite{Allison2016}. In practice, many works in the literature apply SWIR satellite imagery for the detection of wildfires. Some such studies are as follows: Ref.~\cite{Thangavel2023} uses CNNs to classify wildfires with a case study involving Australian wildfires, Ref.~\cite{Barbosa2002} fuses SWIR images with other wavelengths to improve the mapping of burned areas applied to MODIS data in Southern Europe, and Ref.~\cite{Dennison2009} evaluates the effective wavelengths for daytime wildfire detection while considering the impacts of atmospheric attenuation. Additionally, Ref.~\cite{Dennison2009} determined that SWIR bands centered at \SI{2.06}{\upmu\meter} and \SI{2.43}{\upmu\meter} were the most effective for the applied dataset.

\subsection{Planned Constellations for Wildfire Detection} \label{subsec:Lit_Constellations}

There are currently planned constellations of EO satellites dedicated to wildfire detection with purpose-built sensor packages and machine learning. Firstly, Google's FireSat---partnered with Muon Space, the Earth Fire Alliance, and other associated foundations and wildfire authorities---is a planned constellation of $50$ LEO satellites capable of a $\SI{5}{\meter}\times\SI{5}{\meter}$ pixel resolution with a global revisit time of \SI{20}{minutes} \cite{GoogleFireSat,EFAFireSat,Sohel2025WildfireRisk}. FireSat launched a single prototype satellite in March of 2025 \cite{GoogleFireSat}, with three planned to launch in 2026 equipped with multispectral IR instruments containing six bands, including SWIR bands \cite{Honary2025EarlyDetection}. Additionally, FireSat systems utilize machine-learning-based artificial intelligence to compare obtained imagery to the previous $1000$ obtained images of the same location to determine wildfire presence \cite{GoogleFireSat}. Secondly, the WildFireSat constellation is planned to consist of seven LEO microsatellites \cite{WildFireSatCanadaGov} equipped with the Canadian Wildland Fire Monitoring Sensor. capable of observing in four bands, including medium-wave IR (MWIR), with a spatial resolution of \SI{200}{\meter} \cite{WildFireSateoPortal}. The WildFireSat constellation is planned for launch in 2029 and is being developed by the Canadian Space Agency, Natural Resources Canada, and Environment and Climate Change Canada \cite{WildFireSatCanadaGov,Hope2024WildFireSat}. Lastly, the OroraTech Wildfire Constellation presently includes eight 8U CubeSats of a total planned $100$-satellite constellation with a detection resolution of $\SI{4}{\meter}\times\SI{4}{\meter}$ \cite{OroraTecheoPortal}. The sensor onboard each satellite ranges from MWIR to long-wave IR to detect high-temperature and low-temperature flames without impairment from smoke presence \cite{OroraTecheoPortal,Schottl2024OroraTech}. Additionally, each satellite is equipped with onboard machine learning processing capabilities that allow for on-orbit fire detection \cite{Schottl2024OroraTech}. It should be noted that each of these planned constellations is comprised of either nadir-directional or agile systems, with no capability to perform regular maneuvers for a more rapid response, which is a limitation we seek to demonstrate the capability to overcome through the use of constellation reconfigurability.

\subsection{Prior Integration Efforts for Detection and Scheduling} \label{subsec:Lit_Pipeline}

Initial efforts to automate event detection and satellite scheduling have been investigated in prior research and mission design. For instance, the Autonomous Science Agent (ASE) onboard the Earth Observing One (EO-1) spacecraft was an autonomous software component capable of event (and change) detection, including volcanic eruptions and flooding (among others), and schedule modification using the Continuous Activity Scheduling Planning Execution and Replanning (CASPER) software \cite{Chien2004TheEO1}. The operating procedure of the ASE system is summarized as follows: 1) generate a schedule using CASPER relative to targets uplinked from a ground station, 2) analyze acquired imagery from the initial schedule to detect changes, 3) generate a new schedule using CASPER based on detections that includes downlinking detections, and 4) repeat for subsequent observations continuously \cite{Chien2004TheEO1}. Reference~\cite{Davies2006ASE} details a use case of the ASE system to monitor active volcanic sites using the EO-1 Hyperion sensor package, which produces imagery across $196$ bands ranging from \SI{0.4}{\upmu\meter} to \SI{2.5}{\upmu\meter} \cite{Davies2010Hyperion}. Separately from the ASE is the concept of a sensor web, in which a combination of information from multiple independent sensors is utilized cooperatively and, in the cases of Refs.~\cite{Chien2011FireSensorweb,Kangaslahti2024Sensorweb,Chien2020Automated}, is used to autonomously task a constellation of satellites. For instance, Ref.~\cite{Chien2011FireSensorweb} utilizes MODIS, Landsat-5, and EO-1 Advanced Land Imager data to categorize detected wildfires before tasking EO-1 through CASPER with the highest-priority wildfire detections, while Ref.~\cite{Kangaslahti2024Sensorweb} utilizes MODIS, VIIRS, and Synthetic Aperture Radar (from Sentinel-1) data to detect flooding before using the SkySat constellation operated by Planet to obtain subsequent images of regions with detected flooding \cite{Augenstein2016}, and Ref.~\cite{Chien2020Automated} leverages various satellite and ground-based sensor technologies to detect new volcanic activity before tasking EO-1 in the same manner as Ref.~\cite{Chien2011FireSensorweb}.

While these initial efforts operated (or operate) well, each contains some key limitations that more recent advances have attempted to alleviate. First, the details of Ref.~\cite{Davies2006ASE} reveal that ASE relies upon algorithmic machine-learning analysis of Hyperion imagery for the detection of volcanic activity \cite{Burl1998Volcanoes}, while Refs.~\cite{Chien2011FireSensorweb,Kangaslahti2024Sensorweb,Chien2020Automated} additionally rely on similar algorithms for the relevant mission purpose, which may be improved through the use of newly improved CNN architectures. Second, Ref.~\cite{Davies2006ASE} also reveals that CASPER generates a high-level schedule for the duration of one week and a low-level schedule for the duration of one day \cite{Chien2000CASPER} while reordering some tasks within the low-level schedule as necessary \cite{Chien2004TheEO1}. Finally, the constellation tasking request method for the sensor-web approaches in Refs.~\cite{Chien2011FireSensorweb,Kangaslahti2024Sensorweb,Chien2020Automated} result in a long wait time between tasking and data return. In the case of Ref.~\cite{Kangaslahti2024Sensorweb}, this wait is between one and five days due to the lack of ownership over the SkySat constellation, while in the cases of Refs.~\cite{Chien2011FireSensorweb,Chien2020Automated}, the wait is between one and two days \cite{Davies2010Hyperion} due to the uplink windows with EO-1. This wait time can be alleviated through the use of onboard detection and scheduling rather than requiring uplink from the sensor web.

In response to recent advances in object detection in satellite imagery and the growing need for rapid information collection for natural disasters, recent research has begun expanding to combine these newly developed methods back into the pipeline. For instance, CNNs have been developed to process satellite images in orbit, allowing the downlink of only relevant data to ground stations, thereby conserving satellite resources and removing the need for human interpretation \cite{Heimbach2025Managing}. Additionally, Ref.~\cite{Heimbach2025Managing} mentions future work to automate the process of satellite tasking, although currently this has not yet been achieved, leaving a gap between image processing and task scheduling. Secondly, the scheduling of subsequent volcano imaging in Ref.~\cite{Chien2020Automated} is performed using an agile satellite scheduler relying on heuristic methods, which may be improved through the use of both constellation reconfigurability and MILP, both of which reside in the \REOSSP. Finally, relating to wildfire detection, a combination of satellite imagery and manned aerial drones has been used to assess active wildfire fronts and burn maps in near real time \cite{Jimenez2026Leveraging}; however, the satellite systems employed in Ref.~\cite{Jimenez2026Leveraging} are used primarily before and after a wildfire occurs to identify risk and assess the aftermath, respectively. As such, the manned aerial drones fill the role of near-real-time information processing, while the use of satellite imagery may be further improved with autonomous tasking and near-real-time image processing. 

These recent literature developments have provided breakthroughs toward the combination of object detection and autonomous satellite scheduling, while currently addressing fragments of the overall detection and scheduling pipeline or inheriting limitations that could otherwise be alleviated. As such, the goal of this paper is to further these endeavors, improving upon near-real-time wildfire detection by employing CNN architectures and improving upon autonomous satellite tasking through the use of constellation reconfigurability and MILP within the \REOSSP.

\section{WildFire-applicable Intelligent and Responsive Ensemble for Detecting and Scheduling} \label{sec:WildFIRE-DS}

The WildFIRE-DS employs three key modules to enable operations; these include the \textit{Detection Module}, which is used to detect active wildfires within satellite remote sensing images; the \textit{Multi-Pass Confidence Module}, which updates the interpretation confidence obtained by the Detection Module on subsequent flyovers of a potentially detected active wildfire; and the \textit{Schedule Module}, which schedules satellite operations relative to detected active wildfires. Within the automated detection and scheduling process depicted in Fig.~\ref{fig:Illustrative_Figure}, the WildFIRE-DS discretizes a given mission horizon into smaller sections, referred to as Blocks; each Block encompasses the duration of a single schedule within the Schedule Module. In this paper, the Detection Module leverages CNNs with various sensor fusion techniques, the Multi-Pass Confidence Module uses Bayesian statistics to update CNN interpretation confidence values with an impact from prior interpretations, and the Schedule Module employs the \REOSSP from Refs.~\cite{Pearl2025Developing,Pearl2025REOSSP} to provide greater responsiveness through satellite maneuvering capabilities, as well as a baseline EOSSP. As such, each Block contains a given number of stages for constellation reconfiguration with a maximum maneuver budget. Within each Block, four distinct Phases occur, in which various modules are activated:

\begin{enumerate}[label=(Phase \arabic*), leftmargin=1.75cm]
    \item Perform passive observations and execute the Schedule Module's resultant \REOSSP or EOSSP schedule from the previous Block information (where applicable). The purpose of passive observations is to collect satellite images for the Detection Module in Phase~2. During the first Block of a given mission, each satellite performs no orbital maneuvers, as there is no previous Block information with which to determine optimal maneuvering. 
    \item Process each passive observation using the Detection Module such that potential active wildfires at each passive observation location are determined. The Detection Module in this paper leverages a YOLO CNN algorithm, YOLOv11 \cite{Redmon2016YOLO}, with sensor fusion techniques to detect active wildfires with object classification within specified spectral bands. The spectral bands, sensor fusion techniques, and YOLOv11 structure utilized in the Detection Module are explained at length in Sec.~\ref{subsec:CNN}. If one or more active wildfires are detected, the latitude and longitude of the detected active wildfires are calculated relative to the CNN bounding-box center as described in Appendix~A. The detected active wildfire latitude, longitude, and confidence values are provided to Phase~3. 
    \item Update stored active wildfire information relating to all potential detections in Phase~2. Specifically, append all detected active wildfire latitude, longitude, and confidence values to an \textit{auxiliary target set} if they have not been previously added; otherwise, update the relevant auxiliary target set entry according to the Multi-Pass Confidence Module. In this paper, the Multi-Pass Confidence Module updates auxiliary targets using Bayesian statistics for updating recursive probabilities, a process described in more detail in Sec.~\ref{subsec:Multipass}. Upon the conclusion of updating potential active wildfire confidence values in the auxiliary target set, any with a confidence value greater than a threshold of \SI{95}{\%} is added to a \textit{priority target set}. The priority target set is provided to the \REOSSP and EOSSP for scheduling of observations, as it contains targets with a sufficiently high potential to be correctly identified as wildfires. Simultaneously, the $50$ potential active wildfires with the highest confidence values in the auxiliary target set are provided to the \REOSSP for the consideration of more frequent flyovers within the orbital maneuver scheduling. Finally, if the priority target set is expanded during Phase~3, the new priority target set is provided to the EOSSP to update the optimal schedule for the remainder of the Block, relative to the already scheduled orbital maneuvers from the \REOSSP.  
    \item Utilize the updated priority target set, auxiliary target set, and other associated satellite information, such as orbital position and previous data and battery storage values, to optimize the \REOSSP schedule of the subsequent Block. The \REOSSP accounts for information obtained in Phase~3 to schedule satellite tasks of orbital maneuvers, active wildfire observation, data downlink, and solar charging while maintaining operational limits of onboard data and battery storage. The \REOSSP is explained at length in Sec.~\ref{subsec:REOSSP}. 
\end{enumerate}

In the development of the WildFIRE-DS algorithm as a proof of concept of the proposed framework, several key assumptions are made. Firstly, it is assumed that Phases~1 and~2 occur onboard each satellite individually, where the CNN image interpretation results of Phase~2 are then distributed to an autonomous decision-maker, while Phases~3 and~4 occur subsequently within the decision-maker, as indicated in Fig.~\ref{fig:Illustrative_Figure}. To avoid loss of generality, the decision-maker may be hosted by a ground station, a network of ground stations, or a supporting constellation with cross-communication between satellite assets. Secondly, it is assumed that the decision-maker is connected globally and instantaneously with the satellite constellation performing wildfire detection and monitoring.

The process of Block scheduling, priority and auxiliary target set updating, and subsequent Block scheduling repeats until the end of the overall mission horizon, wherein several metrics are obtained to evaluate the performance of the WildFIRE-DS. The schedule of each Block indicates the observation of high-potential targets in the priority target set and the downlink of collected data, thereby presenting both the priority target set for comparison with ground truth and, consequently, the amount of relevant data collected. Additionally, the passive observations performed in Phase~1 allow for the continual updating of both the priority and auxiliary target sets such that newly started wildfires may be tracked for future observation in near real time. Finally, the flexibility provided by constellation reconfigurability allows for the collection of more overall data relating to high-potential targets within the priority target set.

\subsection{Detection Module: Convolutional Neural Network} \label{subsec:CNN}

\begin{figure}[!ht]
    \centering
    \includegraphics[width = 0.65\textwidth]{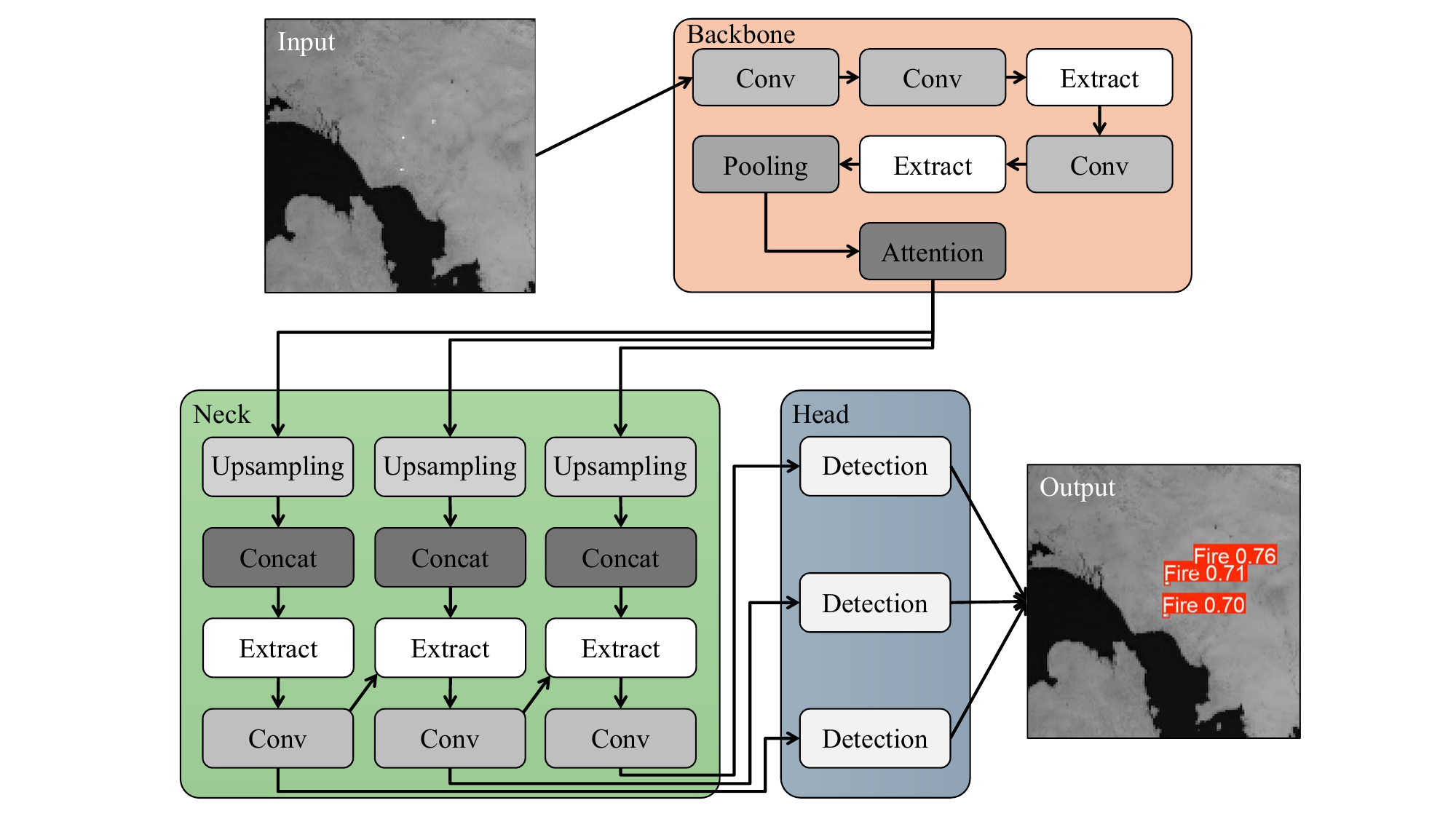}
    \caption{A simplified architecture flow diagram of YOLOv11.}
    \label{fig:YOLO}
\end{figure}

YOLOv11 is selected as the object detection algorithm within the Detection Module of the WildFIRE-DS to handle the complex task of autonomous wildfire detection in satellite imagery. YOLOv11 provides rapid bounding-box detection with classification certainty, enabling accurate and computationally efficient predictions on large datasets \cite{Ali2024TheYOLO}. A simplified visualization of the default YOLOv11 architecture is shown in Fig.~\ref{fig:YOLO}, comprised of three main sections: the Backbone, Neck, and Head \cite{He2025}. The input image is first fed into the Backbone layer, wherein the main feature extraction occurs. The features are continuously enhanced through several operations, including convolution, extraction, pooling, and attention. The processed features are then passed to the Neck layer, in which the multiple enhanced features from the Backbone are clustered into larger enhanced feature representations. This is done through interconnected upsampling, concatenation, extraction, and convolution operations. Lastly, these larger enhanced features are passed to the Head layer, where object classification and bounding-box creation are conducted. After processing the image features, the final output is generated, which includes the object classification, bounding-box coordinates on the image, and the CNN model prediction confidence. 

In this paper, YOLOv11 is trained and validated on a custom dataset of wildfire imagery to suitable levels of accuracy and precision. For this work, the Systems Tool Kit (STK) version 12.10.0, maintained by Ansys Inc. \cite{STK}, is used to generate simulated satellite imagery of wildfires through the Electro-Optical Infrared (EOIR) sensor package. The STK EOIR sensor type can simulate satellite imagery within a custom spectral-band range and associated optical or radiometric properties, such as the field of view and effective focal length. To maintain similarity to works listed in Sec.~\ref{subsec:Lit_SWIR} and ensure realism, the SWIR wavelengths of \SI{1.57}{}--\SI{1.65}{\upmu\meter} and \SI{2.11}{}--\SI{2.29}{\upmu\meter}, corresponding to Bands $6$ and $7$ of the Operational Land Imager system onboard Landsat-8 \cite{Schroeder2016}, are selected to simulate the satellite imagery. Each wavelength band is output as a separate single-channel image with a bit depth of $24$. As a result of the capability of SWIR bands to penetrate smoke and haze \cite{Allison2016}, cloud cover is neglected when simulating satellite imagery. Similarly, horizontal and vertical half-angles of \SI{22.5}{deg} are selected to maintain a \SI{45}{deg} total field of view (FOV) with $128$ pixels in each direction. Each image is generated using a nadir-pointing LEO satellite in a circular orbit positioned directly above land, where the orbital characteristics of the satellite vary randomly; the altitude ranges between \SI{700}{km} and \SI{900}{km}, the inclination ranges between \SI{45}{deg} and \SI{80}{deg}, and the right ascension of the ascending node (RAAN) and argument of latitude range between \SI{0}{deg} and \SI{360}{deg}. Each image is generated at 16:00:00 Coordinated Universal Time (UTC) on August 20th, 2025, such that some images are generated on the day side of the Earth while others are on the night side. An example generated image is shown as the input image in the top left of Fig.~\ref{fig:YOLO}, where the satellite is along the coast of northern France, and there are three visible fires denoted by groupings of bright pixels. 

A simulated dataset curated from the STK EOIR sensor package was deemed appropriate for this paper as opposed to obtaining real-world satellite imagery for a few key reasons. Firstly, real-world satellite imagery is only obtainable from satellite systems that are either nadir-directional or agile within a static orbital trajectory. This aspect introduces an inability to link direct historical geographic-location-associated imagery from exact satellite positions with a maneuverable satellite system that may have altered its orbital trajectory to a location removed from the direct image location. As such, real-world imagery cannot be directly associated with simulated maneuverable satellite systems. Secondly, the STK EOIR sensor package is a high-fidelity simulation environment that can integrate orbit propagation and remote sensing to generate realistic spectral data adequate to very nearly recreate the conditions in which real-world imagery is obtained. Finally, the imagery generated utilizing the STK EOIR sensor package, as shown in Fig.~\ref{fig:YOLO}, has relative similarity to the images generated by the system developed in Ref.~\cite{Fukuhara2017Detection}, at a lower resolution.

\begin{figure}
    \centering
    \includegraphics[width=0.5\textwidth]{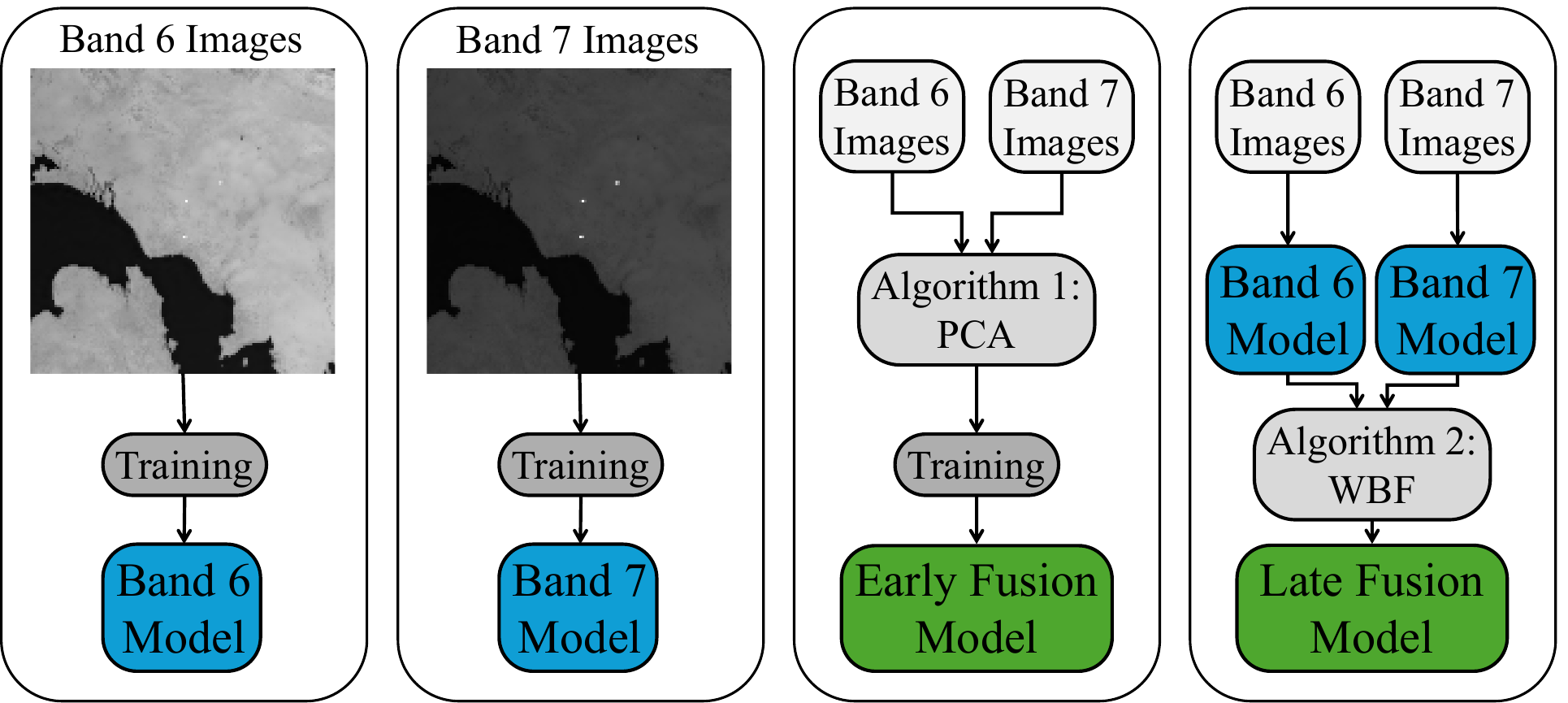}
    \caption{A depiction of the CNN model compositions in this paper.}
    \label{fig:CNN_Models}
\end{figure}

A total of $4000$ satellite images are generated for the training and validation process of four distinct YOLOv11 configurations for comparison, of which $2000$ are generated using Band $6$ and $2000$ are generated using Band $7$, and a further $2000$ are created by fusing the individual Band $6$ and Band $7$ images. The configurations consist of the \textit{Band $6$ Model} trained on a dataset of $1000$ Band $6$ images, the \textit{Band $7$ Model} trained on a dataset of $1000$ Band $7$ images, the \textit{Early Fusion Model} trained on a dataset of $1000$ images created by fusing the Band $6$ and Band $7$ images, and the \textit{Late Fusion Model} wherein the output of the Band $6$ Model and Band $7$ Model are fused. The intricacies of early and late fusion in the Early Fusion Model and the Late Fusion Model are subsequently detailed. The remaining $1000$ images of each Band and $1000$ created early-fusion images are utilized as an additional dataset for the purpose of validating each YOLOv11 model. Each image contains between one and five wildfires in sizes ranging from $1000$ to $5000$ acres, and each wildfire is created as a Gaussian blob using a temperature map within the satellite sensor FOV. More fine-resolution wildfire propagation is not considered in this paper due to the selection of orbital altitude and image pixel resolution, but in general, the incorporation of fine-resolution wildfire propagation would operate in the same manner in the Detection Module. As a result of the wildfire size selected, the average bounding-box size of ground-truth wildfires is $43.35$ pixels squared. Figure~\ref{fig:CNN_Models} depicts each CNN model alongside the structure to obtain each model, including the input images of Band $6$ and Band $7$, the early-fusion algorithm used in the Early Fusion Model to generate fused images, and the late-fusion algorithm used in the Late Fusion Model to combine the outputs of the Band $6$ Model and Band $7$ Model. As a result of the assumption that Phases~1 and~2 occur onboard each satellite individually, it is therefore assumed that the fusion and CNN interpretation processes take place onboard each satellite before potential detections are transferred to the decision-maker.

\subsubsection{Early Fusion}

In the Early Fusion Model, raw data from multiple images is concentrated or combined before undergoing CNN inference. Specifically, the Early Fusion Model employs principal component analysis (PCA) to fuse these inputs, as studies of image fusion suggest that PCA is ideal for fusion quality \cite{Sahu2012Different}. PCA finds the amount of correlated variables in the images and transforms them into uncorrelated variables known as principal components \cite{Naidu2008}. These components are ranked by the variance they represent, with the first component capturing the most significant patterns of the data. By performing eigen-decomposition on the image data covariance matrix, the first principal component is calculated and used to find the normalized weights. These weights are then used to generate a single-weighted fused image, which is finally fed into the CNN for inference. Algorithm~\ref{Alg:Early_Fusion} shows the pseudo-code for this approach.  

\begin{algorithm}[!ht]
    \DontPrintSemicolon
    \caption{Early Fusion.}
    \label{Alg:Early_Fusion}
    Access pixel values for $F$ input satellite images, each with identical height $h$ and width $w$. \;
    Initialize matrix $M$ to store flattened image vectors from $F$ images. \;
    \For{each image $f = 1, \ldots, F$}{
        The image matrix $I^f$ is accessed. \;
        Flatten $I^f$ by appending each row to the previous to obtain vector $\bm{v}^f$ of length $hw$. \
        Append $\bm{v}^f$ as a row to the matrix $M$. \;
        }
    Once the matrix $M$ is constructed with dimensions $F$ by $hw$: \;
    Calculate the covariance matrix $C$ of matrix $M$. \;
    Calculate the eigenvalues and eigenvectors of $C$. \;
    Select the principal component vector $P_{\text{C}}$ as the eigenvector corresponding to the $\arg\max$ of the eigenvalues. \;
    \For{each image index $f=1, \ldots, F$}{
        Calculate the fusion weight $w^f$ based on the principal component $P^f_{\text{C}}$ via: $w^f \gets \frac{P^f_{\text{C}}}{\sum_{f=1}^F{P^f_{\text{c}}}}$ \;
        }
    Calculate the flattened fused image vector $\bm{v}_{\text{final}}$ via: \\$\bm{v}_{\text{final}}\gets\sum_{f=1}^F\bm{v}^fw^f$ \; 
    Reshape $\bm{v}_{\text{final}}$ back to dimensions $h$ by $w$ to output the fused image. \;
\end{algorithm}

\subsubsection{Late Fusion}

The Late Fusion Model operates on the predictions from independently trained models (Band $6$ Model and Band $7$ Model) and consolidates them into one final output. This fusion employs a weighted-box-fusion (WBF) implementation, wherein the final bounding-box coordinates and confidence scores are weighted by the confidence of the individual model's confidence values in their predictions \cite{Solovyev2021}. While straightforward, this methodology is widely adopted for fusing sensor data from multiple sensors \cite{Kim2019,Dalahmeh2024}. Identifying similar bounding boxes across agents is done using an intersection-over-union (IoU) threshold, which determines the amount of spatial overlap (if any) between bounding boxes. After determining similar bounding boxes, fusion via a weighted average produces a single fused bounding box and a consolidated final confidence score. Algorithm~\ref{Alg:Late_Fusion} shows the pseudo-code for this approach.

\begin{algorithm}[!ht]
    \DontPrintSemicolon
    \caption{Late Fusion.}
    \label{Alg:Late_Fusion}
    Access location $(x, y)$, width $(w)$, height $(h)$, and confidence $(C)$ information for all $N$ bounding boxes from the interpretations of $F$ CNN models. \;
    Initialize an empty 2-dimensional array $\mathcal{N}$ where each entry is a bounding box, each column represents a unique bounding box, and each row contains bounding boxes similar to those in the first row of the respective column. \;
    \For{each bounding box $n = 1, \ldots, N$}{
        The information $(x_n, y_n, w_n, h_n, C_n)$ of bounding box $n$ is accessed. \;
        \For{each column $\ell$ of $\mathcal{N}$}{
            The information $(x_{n^\prime}, y_{n^\prime}, w_{n^\prime}, h_{n^\prime}, C_{n^\prime})$ of bounding box $n^\prime$ is accessed from the first row of column $\ell$ in $\mathcal{N}$. \;
            Find the IoU ratio between bounding boxes $n$ and $n^\prime$ using the information $\{x_{n, n^\prime}, y_{n, n^\prime}, w_{n, n^\prime}, h_{n, n^\prime}\}$. \;
            \If{the IoU ratio is greater than an assigned threshold}{
                Bounding box $n$ is similar to bounding box $n^\prime$, append to column $\ell$ of $\mathcal{N}$. \;
                }
            \Else{
                Bounding box $n$ is not similar to bounding box $n^\prime$, append to row one of $\mathcal{N}$. \;
                }
            }
        }
    Once all $N$ bounding boxes are processed, there are $L$ unique bounding boxes (columns of $\mathcal{N}$): \;
    \For{each column $\ell = 1, \ldots, L$ of $\mathcal{N}$}{
        Assign $a = 1, \ldots, A$ as the bounding boxes (non-empty rows) in column $\ell$. \;
        Assign location $(x_\ell, y_\ell)$ and confidence $(C_\ell)$ as follows:\;
        $x_{\ell} \gets \left(\sum_{a = 1}^A x_aC_a\right)/\left(\sum_{a = 1}^A C_a\right)$ \;
        $y_{\ell} \gets \left(\sum_{a = 1}^A y_aC_a\right)/\left(\sum_{a = 1}^A C_a\right)$ \;
        $C_{\ell} \gets \left(\min(A, F)\sum_{a = 1}^A C_a\right)/\left(AF\right)$ \;
        }
\end{algorithm}

\subsubsection{Comparative Analysis}

To determine the effectiveness of each CNN model, a comparative analysis is conducted regarding evaluation metrics from training and model response to the set of $1000$ validation images. The evaluation metrics include the Precision, $P_{\text{r}}$, which measures the proportion of correctly classified wildfires relative to all predicted wildfires, indicating the false alarm rate; the Recall, $R$, which represents the proportion of ground truth wildfires correctly identified, determining if most wildfires were detected; and the F-score, $F$, the harmonic mean of Precision and Recall, presenting a measure of overall detection performance. Additionally, the highest-performing model can be identified using a combined metric known as the mean average precision, mAP, which is determined throughout the training process. The mAP is measured as the combined area of Precision and Recall in a given training epoch (a complete iteration through the full set of $1000$ training images), representing a metric of the tradeoff between Precision and Recall to maximize the effect of each, and is known to be the most informative evaluation metric for object detection algorithms \cite{Wang2022}. The evaluation metrics of Precision, Recall, and F-score are defined by Eqs.~\eqref{eqn:Precision},~\eqref{eqn:Recall}, and~\eqref{eqn:F-Score}, respectively. 
\begin{subequations}
    \begin{align}
        P_{\text{r}} &= \frac{\text{true positive}}{\text{true positive} + \text{false positive}}
        \label{eqn:Precision} \\
        R &= \frac{\text{true positive}}{\text{true positive} + \text{false negative}}
        \label{eqn:Recall} \\
        F &= 2\left(\frac{P_{\text{r}} R}{P_{\text{r}} + R}\right)
        \label{eqn:F-Score}
    \end{align}
\end{subequations} 

\begin{figure}[!ht]
    \centering
    \includegraphics[width = \textwidth]{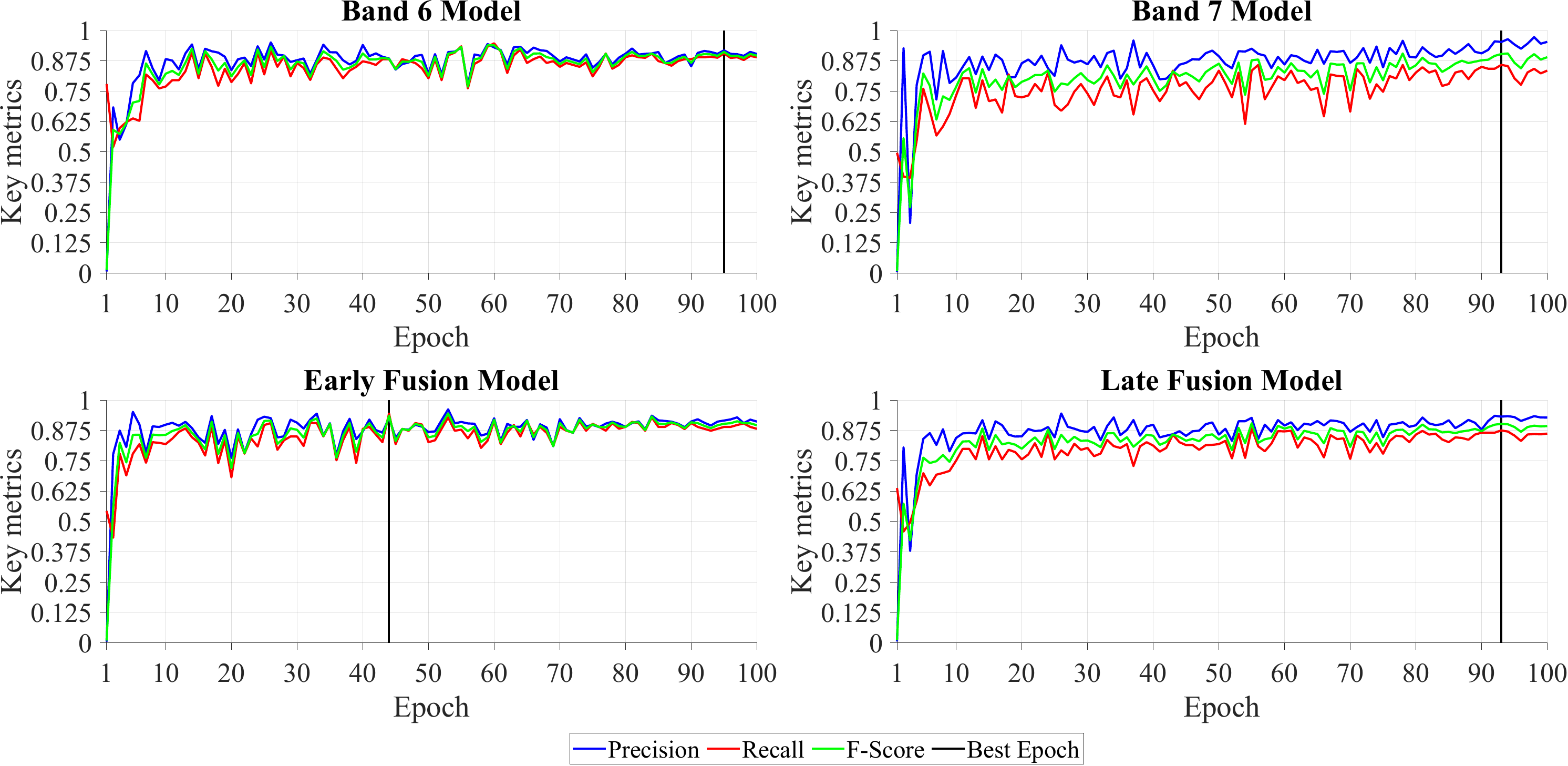}
    \caption{CNN training metrics over $100$ Epochs.}
    \label{fig:CNN_Model_Metrics}
\end{figure}

Using these evaluation metrics along with the highest mAP, each model exhibits somewhat favorable results after $100$ epochs. Figure~\ref{fig:CNN_Model_Metrics} shows the evaluation metrics over all $100$ training epochs for all four models, with a vertical line indicating the epoch with the highest mAP. Overall, each model demonstrates a steep performance improvement during early training, stabilizing after approximately $10$ epochs. The plateau in metric improvement over the final $90$ epochs implies that each model converged for the given set of training images. Visually, the Band $6$ Model and Early Fusion Model converged more consistently than the Band $7$ Model and Late Fusion Model. Upon closer analysis of the training metrics, the highest mAP values for the Band $6$ Model, Band $7$ Model, Early Fusion Model, and Late Fusion Model are \SI{70.43}{}, \SI{65.47}{}, \SI{69.14}{}, and \SI{67.26}{\%} at epoch $95$, $93$, $93$, and $44$, respectively. Furthermore, Table~\ref{tab:CNN_Model_Metrics} displays the evaluation metrics of Precision, Recall, and F-score at the highest performing epoch for each model, demonstrating that the Early Fusion Model has the highest Recall and F-score values, while the Band $7$ Model has the highest Precision value. As such, the evaluation metrics during training suggest that the Early Fusion Model is the highest performing CNN model, which will be further investigated through each model's response to the set of validation images. 

\begin{table}[!ht]
    \centering
    \caption{CNN training metrics at the highest performing epoch.}
    \begin{tabular}{ l r r r r }
\hline \hline
Metric        & Band $6$ Model & Band $7$ Model & Early Fusion Model & Late Fusion Model \\
\hline
Precision, \% & $91.81$  & $95.26$  & $92.78$      & $93.53$     \\
Recall, \%    & $90.55$  & $85.83$  & $94.49$      & $88.19$     \\
F-score, \%   & $91.18$  & $90.30$  & $93.63$      & $90.74$     \\
\hline \hline
    \end{tabular}
    \label{tab:CNN_Model_Metrics}
\end{table}

\begin{figure}[!ht]
    \centering
    \includegraphics[width = 0.75\textwidth]{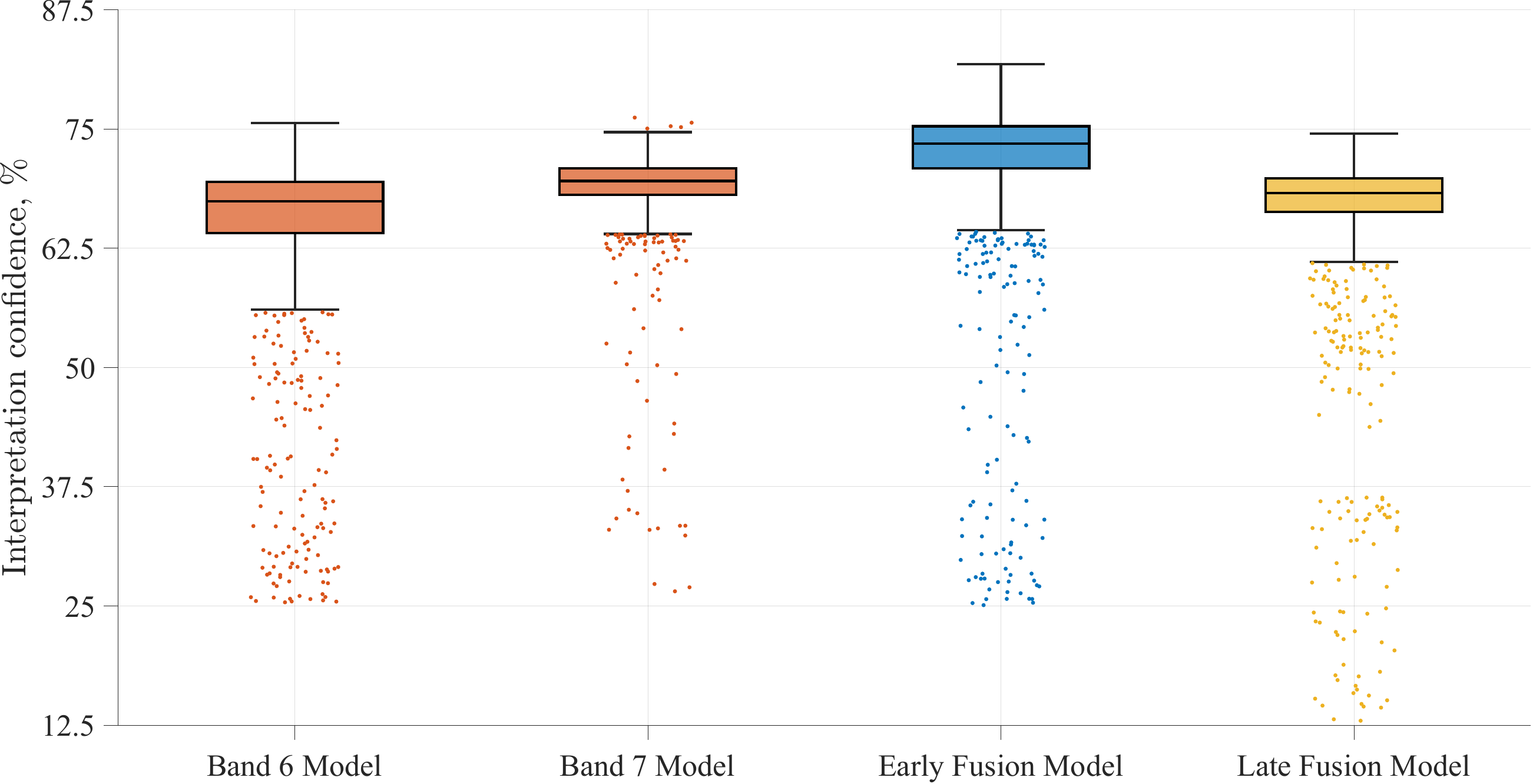}
    \caption{CNN model responses to $1000$ validation images.}
    \label{fig:CNN_Validation_Response}
\end{figure}

Additionally, the response of each model to the set of $1000$ validation images provides further insight into the performance of each model. Firstly, Table~\ref{tab:CNN_Model_Metrics_Val} displays the evaluation metrics of Precision, Recall, and F-score for each model in response to the validation datasets, wherein the Band $7$ Model has the highest values in all three metrics, while the Band $6$ Model has the lowest values in all three metrics. As a result, the Late Fusion Model is second best in all three metrics, as it relies on a combination of the Band $6$ Model and Band $7$ Model for operation, and the Early Fusion Model is the second worst in all three metrics. The differences in metrics between each model’s performance on the training and validation datasets arise from the stochasticity introduced during training by the default YOLOv11 architecture, which is disabled during validation, resulting in slightly higher validation metrics. Secondly, Figure~\ref{fig:CNN_Validation_Response} displays the CNN model confidence values of each detected wildfire within the response to all validation images as a box chart, where each box represents the inner quartiles relative to the median line within the box, the lines extending beyond the box (whiskers) represent data extending to the outer \SI{10}{\%} bounds, and outliers are marked with dots above or below the whiskers. In parallel, Table~\ref{tab:CNN_Validation_Response} reports $10$ important numeric statistics obtained from Fig.~\ref{fig:CNN_Validation_Response}, including the number of outliers, and lower and upper bounds of various box properties. From Fig.~\ref{fig:CNN_Validation_Response}, the Band $7$ Model has the tightest grouping of data, an observation supported by Table~\ref{tab:CNN_Validation_Response} as the Band $7$ Model has the lowest number of outliers and lowest standard deviation. Additionally, from Fig.~\ref{fig:CNN_Validation_Response}, the Early Fusion Model appears to have the most ideal image interpretation confidence, with higher confidence values. This observation is also supported by Table~\ref{tab:CNN_Validation_Response}, which shows that the Early Fusion Model outperforms each of the CNN models in seven categories. Therefore, despite the evaluation metrics on the validation datasets indicating that the Band $7$ Model is superior to the Early Fusion Model, the interpretation confidence values and training responses strongly support the Early Fusion Model's overall performance.

\begin{table}[!ht]
    \centering
    \caption{CNN validation metrics.}
    \begin{tabular}{ l r r r r }
\hline \hline
Metric        & Band $6$ Model & Band $7$ Model & Early Fusion Model & Late Fusion Model \\
\hline
Precision, \% & $95.43$  & $98.86$  & $95.79$      & $97.14$     \\
Recall, \%    & $97.27$  & $99.38$  & $97.35$      & $98.33$     \\
F-score, \%   & $95.93$  & $99.06$  & $96.29$      & $97.49$     \\
\hline \hline
    \end{tabular}
    \label{tab:CNN_Model_Metrics_Val}
\end{table}

\begin{table}[!ht]
    \centering
    \caption{Statistics of CNN model interpretation confidence values in response to $1000$ validation images.}
    \begin{threeparttable}
        \begin{tabular}{ l r r r r }
\hline \hline
Statistic              & Band $6$ Model & Band $7$ Model & Early Fusion Model & Late Fusion Model \\
\hline
\# Outliers            & 138            & \textbf{84}\tnote{a} & 137            & \textit{167}\tnote{b} \\
Mean, \%               & \textit{65.43} & 69.01                & \textbf{71.88} & 66.51                 \\
Median, \%             & \textit{67.39} & 69.55                & \textbf{73.44} & 68.28                 \\
Standard deviation, \% & 7.46           & \textbf{3.95}        & 7.21           & \textit{7.72}          \\
Minimum, \%            & 25.38          & \textbf{26.53}       & 25.08          & \textit{12.97}        \\
Maximum, \%            & 75.64          & 76.19                & \textbf{81.79} & \textit{74.49}        \\
Lower quartile, \%     & \textit{64.08} & 68.10                & \textbf{70.89} & 66.29                 \\
Upper quartile, \%     & \textit{69.43} & 70.86                & \textbf{75.28} & 69.83                 \\
Min whisker, \%        & \textit{56.07} & 63.99                & \textbf{64.38} & 61.04                 \\
Max whisker, \%        & 75.64          & 74.67                & \textbf{81.79} & \textit{74.49}        \\
\hline \hline
        \end{tabular}
        \begin{tablenotes}
           \item[a] A \textbf{bold} entry denotes the best model for the statistic row
           \item[b] An \textit{italic} entry denotes the worst model for the statistic row
        \end{tablenotes}
    \end{threeparttable}
    \label{tab:CNN_Validation_Response}
\end{table}

As a result of the CNN model training evaluation metrics and response to validation data, the Early Fusion Model is determined to be the highest-performing CNN model investigated in this paper. Therefore, the Early Fusion Model will be prioritized for use in the simulated experimentation conducted in Sec.~\ref{sec:Experiment}. However, the Late Fusion Model is also employed in the simulated experimentation to highlight the modularity of the WildFIRE-DS and provide further comparison between early and late fusion techniques. It should be noted that the WildFIRE-DS assumes that the same CNN model is deployed for operation on each satellite for each instance of experimentation, such that multiple CNN models are not employed in the same experiment instance but can instead be compared across separate instances.

The weights of the Band $6$ Model, Band $7$ Model, Early Fusion Model, and Late Fusion Model, as well as the function code for Algorithms~\ref{Alg:Early_Fusion} and~\ref{Alg:Late_Fusion} are available upon request.

\subsection{Multi-Pass Confidence Module: Bayesian Statistics} \label{subsec:Multipass}

The Multi-Pass Confidence Module in this paper is put in place to incorporate the maximum amount of available information on each wildfire detection, allowing multiple flyovers of a given location to contribute new interpretation data over time. Additionally, because orbital maneuvers are costly, more information about the wildfire is essential to make an informed decision regarding trajectory adjustments. The Multi-Pass Confidence Module employs Bayesian statistics, allowing the desired recursive nature while consistently incorporating information from all previous interpretation confidence values. The principle of Bayesian statistics requires that the same testing method be utilized for each interpretation used in the update process, which holds as a result of deploying the same CNN model across all satellites in each experiment instance. As such, a more informed confidence value is obtained with each new interpretation of previously accounted-for wildfire detections. 

The process used to perform a confidence value update is shown in Eq.~\eqref{Bayes:Updating} relative to an update upon the $i\text{th}$ interpretation of a given potential active wildfire in the auxiliary target set. Given standard Bayesian notation for recursive Bayesian iteration \cite{Gelman2020Bayesian,Puga2015Bayes}, Table~\ref{tab:Bayes} includes the notation and definition of each parameter in Eq.~\eqref{Bayes:Updating}. In Eq.~\eqref{Bayes:Updating}, the left-hand side represents the updated confidence value at interpretation $i$ relative to the hypothesis $H$, that being that an active wildfire is present within the interpretation data. The updated confidence value is computed with respect to the current isolated interpretation confidence value, the previously updated confidence value relative to all previous interpretation data, and the evidence term given in the denominator from the total law of probability. The total law of probability states that the sum of all possible outcomes equals the sum of their probabilities; in this case, it is the sum of the probabilities of a true-positive and a false-positive wildfire interpretation. Therefore, the evidence term includes the current isolated interpretation confidence value, the previously updated confidence value, the accuracy, that being the false-positive rate, of the relevant Detection Module, and the complement of the previously updated confidence value. The first term of the denominator represents the true-positive rate through the inclusion of the current isolated interpretation confidence value and the previously updated confidence value, while the second term of the denominator represents the false-positive rate by providing input of the accuracy of the Detection Module and the complement of the previously updated confidence value. As a result, the inclusion of all relevant information allows the current isolated interpretation confidence value to update the previously updated confidence value, thus allowing the continuous contribution of new interpretation data, refining the interpretation confidence value of potential wildfires over time before they are added to the priority target set. A specific note of the value of $P(D_i \vert \neg H)$ is made in Sec.~\ref{sec:Experiment}.
\begin{equation}
    P(H \vert \{D_i\}) = \frac{P(D_i \vert H) P(H \vert \{D_{i-1}\})}{P(D_i \vert H)P(H \vert \{D_{i-1}\}) + P(D_i \vert \neg H)(1-P(H \vert \{D_{i-1}\}))}
    \label{Bayes:Updating}
\end{equation}

\begin{table}[!ht]
    \centering
    \caption{Notation utilized in Eq.~\eqref{Bayes:Updating}.}
    \resizebox{\textwidth}{!}{
    \begin{tabular}{ l l }
\hline\hline
Notation & Definition \\
\hline
$P(H \vert \{D_i\})$ & Updated confidence value at interpretation $i$ according to the set of all prior interpretation data $\{D_i\}$ \\
$P(H \vert \{D_{i-1}\})$ & Previously updated confidence value from interpretation $i-1$ according to the set of interpretation data $\{D_{i-1}\}$ \\
$P(D_i \vert H)$ & Current isolated interpretation confidence value of interpretation $i$ \\
$P(D_i \vert \neg H)$ & Accuracy/false-positive rate of relevant Detection Module \\
\hline\hline
    \end{tabular}
    }
    \label{tab:Bayes}
\end{table}

\subsection{Schedule Module: The Reconfigurable Earth Observation Satellite Scheduling Problem} \label{subsec:REOSSP}

\begin{figure}[!ht]
    \centering
    \includegraphics[width=0.9\textwidth]{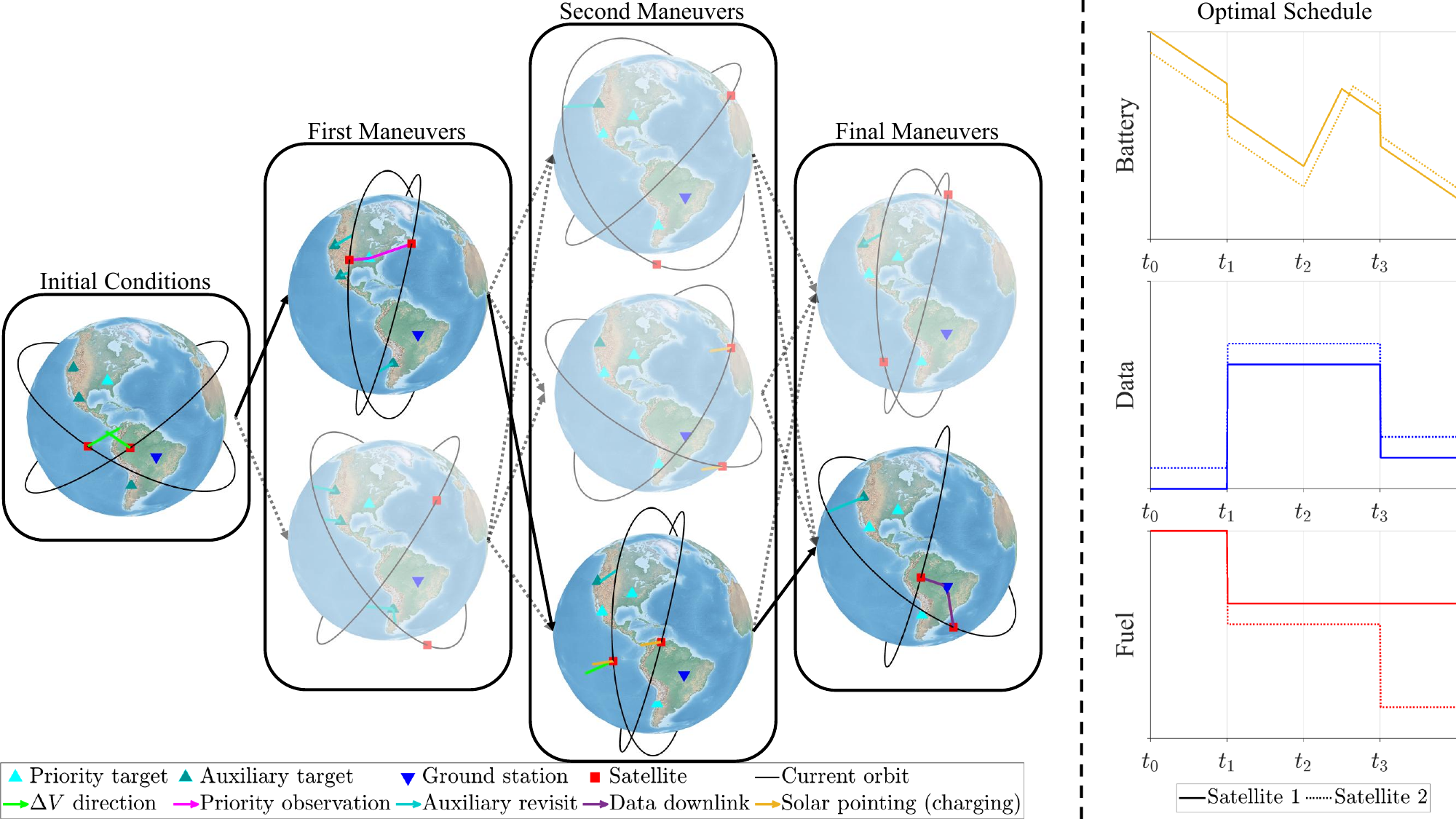}
    \caption{A demonstration of the capabilities in and modifications to the \textcolor{myblue}{\textsf{REOSSP}} \cite{Pearl2025Developing,Pearl2025REOSSP}.}
    \label{fig:REOSSP_Demonstration}
\end{figure}

The Schedule Module leverages a slightly modified version of the \REOSSP developed in Refs.~\cite{Pearl2025REOSSP,Pearl2025Developing}, which optimizes the performance of satellite tasks including orbital maneuvering, priority target observation, auxiliary target revisit, data downlink to ground stations, and solar charging over a fixed schedule duration. The \REOSSP additionally tracks onboard data and power storage levels while obeying a maneuver budget and governing the orbital maneuver sequence according to the Multistage Constellation Reconfiguration Problem \cite{lee2024deterministic}. A demonstration of the capabilities and decision-making process of the \REOSSP is shown in Fig.~\ref{fig:REOSSP_Demonstration} through the high-level depiction of a collection of feasible maneuver options. In Fig.~\ref{fig:REOSSP_Demonstration}, the maneuver sequence that results in the optimal schedule is linked with solid arrows, while maneuvers that result in suboptimal schedules are faded. Within the optimal schedule depicted, priority target observation (indicated through magenta arrows) and auxiliary target revisit (indicated through turquoise arrows) result after optimal maneuvering provides previously unavailable visibility of targets. The optimal schedule then progresses with a priority target set update (indicated through auxiliary targets changing color) and solar charging (indicated through orange arrows) prior to the final maneuvers taking place. Finally, the optimal schedule involves data downlink to ground stations (indicated through purple arrows) and continued auxiliary target revisits. Figure~\ref{fig:REOSSP_Demonstration} demonstrates the overall potential effects of suboptimal maneuvering, wherein the \REOSSP utilizes MILP to search through feasible maneuvers and resulting schedules to find the most effective schedule that maximizes priority target observation, auxiliary target revisit, and data downlink to ground stations.

The fixed finite schedule duration is defined as $T_{\text{r}} \ge 0$ and is discretized by a step size $\Delta t > 0$, resulting in a finite number of time steps $T = T_{\text{r}} / \Delta t$ which are contained in the set of time steps $\mathcal{T} = \{1, 2, \ldots, T\}$. Within the scheduled duration, each satellite may perform orbital maneuvers at $S>0$ equally spaced stages concerning stage $s=0$ as the initial state of the constellation; the set of stages is therefore defined as $\mathcal{S} = \{0, 1, 2, \ldots, S\}$. Since stages are equally spaced, the number of time steps within each stage is $T^s = T/S$ such that the set of time steps in each stage is $\mathcal{T}^s = \{1, 2, \ldots, T^s\}$. Separately, the \REOSSP considers $K > 0$ total satellites within the set $\mathcal{K} = \{1, 2, \ldots, K\}$ which are capable of performing orbital maneuvers between orbital slots within the set of orbital slots $\mathcal{J}^{sk} = \{1, 2, \ldots, J^{sk}\}$ with $J^{sk} \ge 0$ total orbital slots for stage $s \in \mathcal{S}$ and satellite $k \in \mathcal{K}$. Additionally, the \REOSSP considers $P \ge 0$ total priority targets for observation, $P^\prime \ge 0$ auxiliary target for consideration of subsequent revisits, and $G \ge 0$ total ground stations for data downlink within the sets $\mathcal{P} = \{1, 2, \ldots, P\}$, $\mathcal{P}^\prime = \{1, 2, \ldots, P^\prime\}$, and $\mathcal{G} = \{1, 2, \ldots, G\}$, respectively. 

The \REOSSP utilizes various parameters to constrain the problem for physical and operational considerations. Firstly, the visibility of a target $p \in \mathcal{P}$ relative to satellite $k \in \mathcal{K}$ located in orbital slot $j \in \mathcal{J}^{sk}$ relative to stage $s \in \mathcal{S} \setminus \{0\}$ at time $t \in \mathcal{T}^s$ is contained in $V^{sk}_{tjp} \in \{0, 1\}$ where a value of one indicates visibility. Similarly, the visibility of auxiliary target $p \in \mathcal{P}^\prime$, ground station $g \in \mathcal{G}$, and the sun relative to satellite $k$ located in orbital slot $j$ relative to stage $s$ at time $t$ is contained in $U^{sk}_{tjp} \in \{0, 1\}$, $W^{sk}_{tjg} \in \{0, 1\}$, and $H^{sk}_{tj} \in \{0, 1\}$, respectively. Separately, the cost to maneuver from one orbital slot $i \in \mathcal{J}^{s-1, k}$ to another orbital slot $j \in \mathcal{J}^{sk}$ relative to satellite $k$ and stage $s$ is contained in $c^{sk}_{ij} \ge 0$. Furthermore, the maximum maneuver budget for the schedule is applied to each satellite $k$ as $c^k_{\max} \ge 0$. 

To track the data and battery storage levels for each satellite $k \in \mathcal{K}$ from one time step to the next, various parameters are defined. Firstly, the data storage level at each time step is determined relative to the previously conducted target observations and data downlink to ground stations, wherein target observation adds data to the onboard satellite memory, and data downlink to ground stations allows for the deletion of transmitted data. The amounts of data gained through target observation and depleted through data downlink to ground stations are defined as $D_{\text{obs}}$ and $D_{\text{comm}}$, respectively. Additionally, the data storage level must not fall below a minimum threshold or exceed a maximum threshold, defined as $D_{\min}$ and $D_{\max}$, respectively. Similarly, the battery storage level at each time step is determined relative to the previously conducted solar charging, target observation, data downlink to ground stations, orbital maneuvers, and standard operational tasks (such as telemetry and timekeeping); only solar charging accumulates battery while the other task performances require some amount of power to be performed. The battery accumulated or depleted by tasks, in the order previously listed, are defined as $B_{\text{charge}}$, $B_{\text{obs}}$, $B_{\text{comm}}$, $B_{\text{recon}}$, and $B_{\text{time}}$. The battery storage level is also enforced not to fall below a minimum threshold or exceed a maximum threshold, defined as $B_{\min}$ and $B_{\max}$, respectively.

\subsubsection{Decision Variables}

The decision variables and indicator variables of the \REOSSP are the satellite tasks (decision variables) of orbital maneuvers, priority target observation, data downlink to ground stations, and solar charging, as well as the indication of obtained auxiliary target visibility and onboard data and battery storage levels. Each decision variable is binary, such that a value of one indicates the performance of the tasks and a value of zero indicates no task performance. Additionally, the indicator variable for the obtained auxiliary target visibility is binary, such that a value of one indicates obtained visibility, while all other indicator variables are real values. The first decision variable, $x^{sk}_{ij}$, is the ordering of orbital maneuvers of satellite $k$ from an orbital slot $i \in \mathcal{J}^{s-1, k}$ in the previous stage to an orbital slot $j \in \mathcal{J}^{sk}$ in the current stage for all stages $s \in \mathcal{S} \setminus \{0\}$. Additionally,  $\mathcal{J}^{0k}$ is a singleton set representing the initial conditions of satellite $k$. The second decision variable, $y^{sk}_{tp}$, is the observation of priority target $p \in \mathcal{P}$ performed by satellite $k$ at time $t$ in stage $s \in \mathcal{S} \setminus \{0\}$. The third decision variable, $q^{sk}_{tg}$ is the downlink of data to ground station $g$ performed by satellite $k$ at time $t$ in stage $s \in \mathcal{S} \setminus \{0\}$. Finally, the fourth decision variable, $h^{sk}_t$ is the solar charging of satellite $k$ at time $t$ in stage $s \in \mathcal{S} \setminus \{0\}$. Similarly, the indicator variable $\alpha^{sk}_{tp}$ denotes the obtained visibility of auxiliary target $p \in \mathcal{P}^\prime$ by satellite $k$ at time $t$ in stage $s \in \mathcal{S} \setminus \{0\}$. Separately, the indicator variables $d^{sk}_t$ and $b^{sk}_t$ denote the data and battery storage level, respectively, of satellite $k$ at time $t$ in stage $s \in \mathcal{S} \setminus \{0\}$. 

\subsubsection{Full Formulation}

Utilizing the detailed parameters and decision variables, the \REOSSP operates as in Refs.~\cite{Pearl2025REOSSP,Pearl2025Developing} with the exceptions of the following modifications. Firstly, constraint~\eqref{REOSSP:Aux_visibility} applies the auxiliary target visibility $U^{sk}_{tjp}$ according to the currently occupied orbital slot assigned by the decision variable $x^{sk}_{ij}$. Secondly, the objective function is modified to include the consideration of the obtained auxiliary target visibility, thus maximizing and balancing the number of data downlink occurrences, priority target observations, and obtained auxiliary target visibility, as shown in Eq.~\eqref{REOSSP:obj}. Within the objective function, the weighting parameter $C$ is applied to the downlink of data to ground stations, and the weighting parameter $O_p, ~ \forall p \in \mathcal{P}^\prime$ is applied to the obtainment of auxiliary target visibility.
\begin{equation}
    \sum_{i \in \mathcal{J}^{s-1, k}} \sum_{j \in \mathcal{J}^{sk}} U^{sk}_{tjp} x^{sk}_{ij} \ge \alpha^{sk}_{tp}, \qquad \forall s \in \mathcal{S} \setminus \{0\}, \forall t \in \mathcal{T}^s, \forall p \in \mathcal{P}^\prime, \forall k \in \mathcal{K}
    \label{REOSSP:Aux_visibility}
\end{equation}
\hypertarget{REOSSP}{}
\begin{equation} 
    \max \quad z = \sum_{k \in \mathcal{K}} \sum_{s \in \mathcal{S} \setminus \{0\}} \sum_{t \in \mathcal{T}^s} \left( \sum_{g \in \mathcal{G}} C q^{sk}_{tg} + \sum_{p \in \mathcal{P}} y^{sk}_{tp} + \sum_{p \in \mathcal{P}^\prime} O_p\alpha^{sk}_{tp} \right) \label{REOSSP:obj}
\end{equation}

The objective function balances the number of priority target observations, the number of data downlink occurrences to ground stations, and the number of obtained auxiliary target visibility. Concurrently, the constraints restrict feasible schedules such that data is observed before downlink to ground stations, there is available data storage capacity for priority target observation, there is available battery storage to perform operations, and the orbital maneuvering sequence is logical while obeying the maximum maneuver budget. The optimal decision variables returned by the \REOSSP for a given schedule's parameters indicate the scheduled occurrences of priority target observation, data downlink to ground stations, solar charging, maneuver sequencing, auxiliary target visibility windows, and the progression of the onboard data and battery storage levels. The value of the modifications made to the \REOSSP by including auxiliary target set consideration results from incorporating auxiliary target visibility in optimal orbital maneuvers, while simultaneously scheduling operations relative to the priority target set. As such, one can set the weighting parameter $C > 1$, while also setting the weighting parameter $O_p < 1, ~ \forall p \in \mathcal{P}^\prime$, to more heavily favor data downlink and priority target observation while additionally considering auxiliary target visibility when selecting orbital maneuvers.

\subsubsection{The Baseline EOSSP}

This paper also makes use of a baseline EOSSP within the Schedule Module. The baseline EOSSP employed follows the form of the \REOSSP from Ref.~\cite{Pearl2025REOSSP} without modification for the incorporation of auxiliary targets, while also setting the parameter $c^k_{\max} = 0, ~ \forall k \in \mathcal{K}$ to not incorporate constellation reconfigurability. As such, the baseline EOSSP similarly incorporates tasks of priority target observation, data downlink to ground stations, and solar charging, while tracking onboard data and battery storage values, during a given mission duration for a constellation of satellites.

\section{Experimentation} \label{sec:Experiment}

To demonstrate the performance and operation of the WildFIRE-DS algorithm, a computational experiment is conducted with fixed mission parameters. The experimentation includes a multilevel comparison between the \REOSSP and a baseline EOSSP, as well as between the Early Fusion Model and Late Fusion Model employed in the Detection Module. All experimentation is conducted on a platform equipped with an Intel Core i9-12900 $2.4$ GHz (base frequency) CPU processor ($16$ cores and $24$ logical processors) and 64 GB of RAM. Additionally, the EOSSP and \REOSSP are programmed in MATLAB \cite{MATLAB} with the use of YALMIP \cite{YALMIP} and are solved using the commercial software package Gurobi Optimizer (version 13.0.0) with default settings and a runtime limit of one hour for each Block and responsive scheduling. Furthermore, the CNN models are programmed in Python (version 3.9.6) \cite{Python} with the use of the YOLOv11 architecture \cite{Ali2024TheYOLO}, and all satellite imagery for CNN model interpretation is generated within STK \cite{STK}.

The fixed parameters of the computational experiment are as follows: The mission horizon is $14$ days beginning on August $7$\textsuperscript{th}, 2024, at midnight (00:00:00) in Coordinated Universal Time. The mission horizon is discretized into Four Blocks such that the schedule duration $T_{\text{r}}$ of each \REOSSP for each Block is \SI{3.50}{days}. Furthermore, the discrete time step size $\Delta t$ is $100$ seconds, such that $T = 3024$, and the number of stages is $S = 4$, such that each stage of reconfiguration is slightly less than one day. Passive observations are assumed to occur every time step, for a total of \num{12096} passive observations over the entire time horizon for each satellite. Additionally, the satellite orbital data of NOAA-20 and NOAA-21, obtained from Ref.~\cite{Spacetrack}, are used for $K = 2$ satellites. 

The parameters $V$ and $W$ are generated through the use of the MATLAB \texttt{access} function, while $H$ is generated through the use of the MATLAB \texttt{eclipse} function, both of which functions are contained within the Aerospace Toolbox \cite{MATLAB}. Separately, the parameter $O_p,~ \forall p \in \mathcal{P}^\prime$ is set to the confidence value of the respective auxiliary target set squared such that $O_p < 1, ~ \forall p \in \mathcal{P}^\prime$, and the metric $P(D_i \vert \neg H)$ used in the Multi-Pass Confidence Module is the complement of the mAP value of the CNN model in use. As a result of the orbital period of each satellite, passive observations occur roughly $61$ times per orbital period. Each passive observation image is generated in the same manner and with the same instrument parameters in STK \cite{STK} as previously described in the generation of the training data set, but relative to the time of the passive observation occurrence. 

\begin{figure}[!ht]
    \centering
    \begin{subfigure}[h]{0.49\textwidth}
        \centering
        \includegraphics[width = \textwidth]{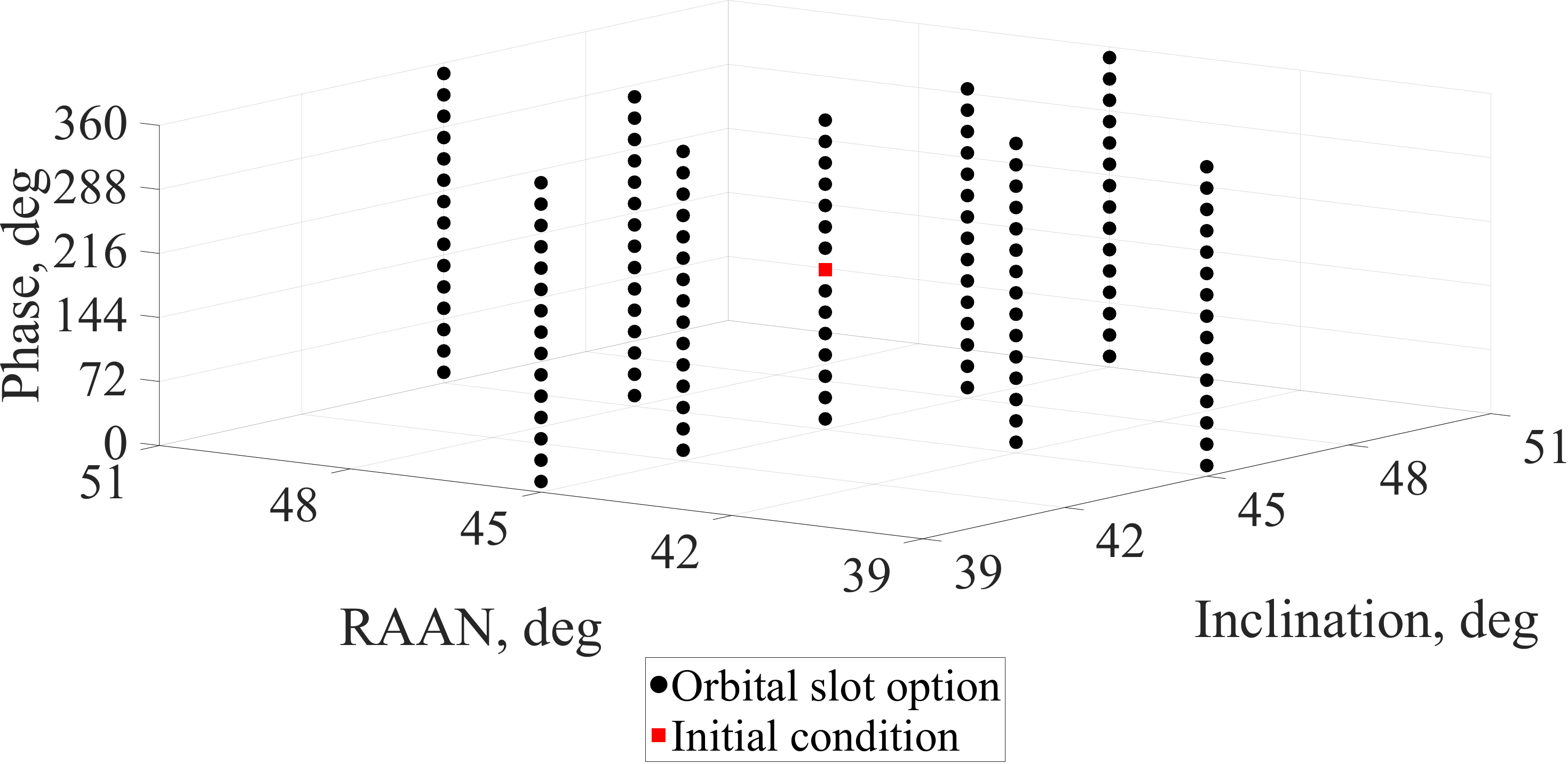}
        \caption{Orbital slot options in $\mathcal{J}^{sk}$.}
        \label{fig:OrbitalSlots}
    \end{subfigure}
    \hfill
    \begin{subfigure}[h]{0.49\textwidth}
        \centering
        \includegraphics[width = \textwidth]{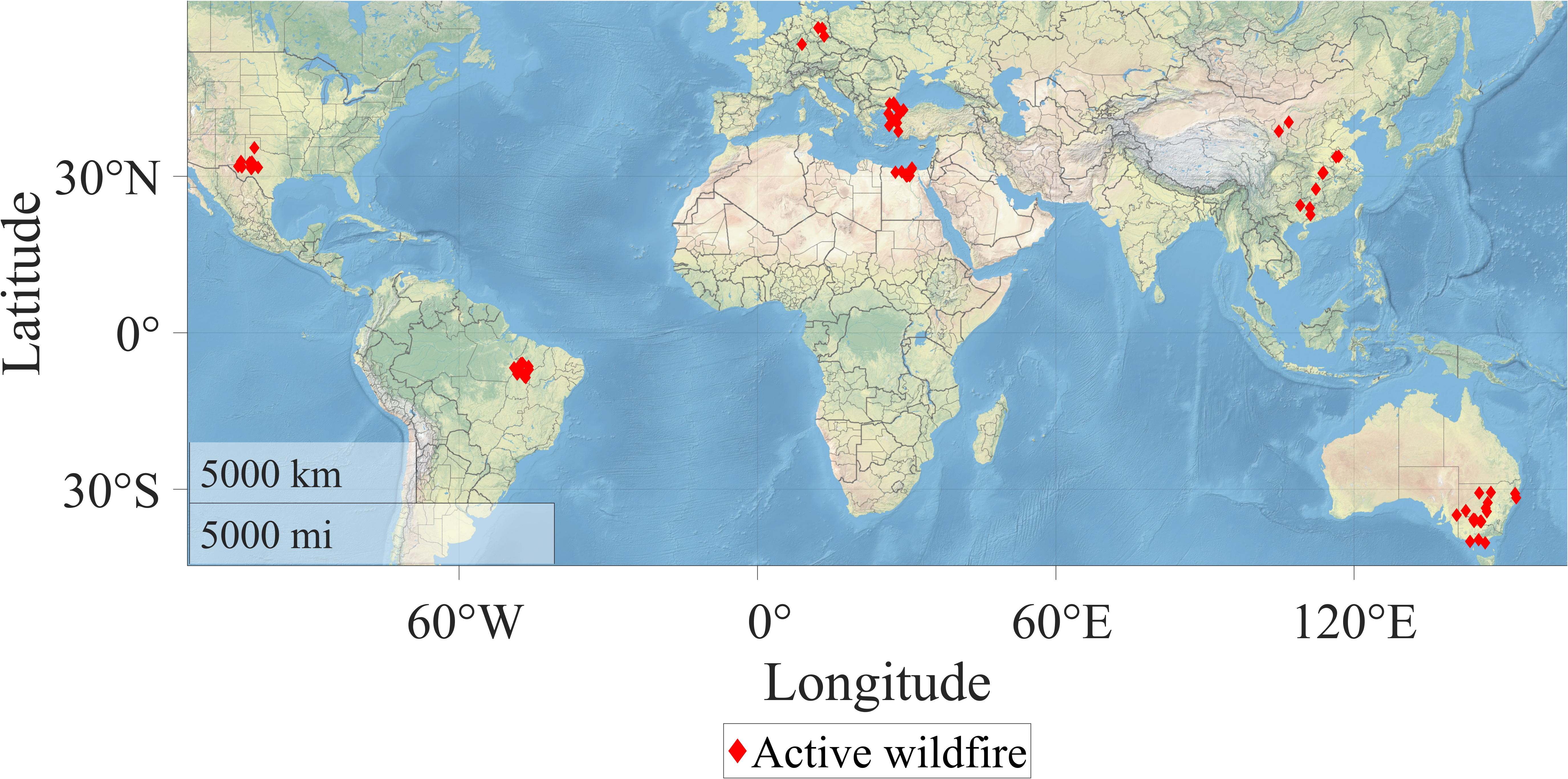}
        \caption{All wildfire events considered for the mission horizon.}
        \label{fig:Wildfires}
    \end{subfigure}
    \caption{Experiment orbital slot options and active wildfire parameters.}
    \label{fig:Slots_and_Fires}
\end{figure}

Each satellite is provided a maximum maneuver budget of $c^k_{\max} = \SI{1.00}{km/s}$ (totaling to \SI{2.00}{km/s} with $K=2$ satellites) with orbital slots that vary in inclination, right ascension of the ascending node (RAAN), and true anomaly, wherein five plane orbital slot options of either type of plane (inclination or RAAN) and $15$ true-anomaly orbital slot options are provided such that $J^{sk} = 135$. The maximum distance between orbital slots is calculated using rearranged analytical expressions from Ref.~\cite{Vallado2022}, assuming the full budget is used. Figure~\ref{fig:OrbitalSlots}, as obtained from Refs.~\cite{Pearl2025Developing,Pearl2025REOSSP}, depicts the available orbital slot option space. The parameter $c$ is computed for each set of orbital slots $i \in \mathcal{J}^{s-1, k}$ and $j \in \mathcal{J}^{sk}$ through various single- or double-impulse orbital transfer maneuvers from Ref.~\cite{Vallado2022}. 

At the start of each Block, the final characteristics of the previous Block, such as the final orbital slot, the final data and battery storage levels, and the amount of the maximum maneuver budget used to perform maneuvers, are used to update the initial conditions of the respective parameter. Moreover, only half of the remaining maximum maneuver budget is available to each Block, apart from the final Block, wherein the entire remaining maximum maneuver budget is available. Additionally, two ground stations are assigned as two used by the Disaster Monitoring Constellation, those being a satellite station in Boecillo Spain at \SI{41.54}{deg} North and \SI{4.70}{deg} West and the Svalbard Satellite Station in Sweden at \SI{78.23}{deg} North and \SI{15.41}{deg} East \cite{DMC_gs,Pearl2025Developing,Pearl2025REOSSP}. Finally, all data storage and battery storage parameters are consistent with those used in Refs.~\cite{Pearl2025Developing,Pearl2025REOSSP}. 

Historical wildfire data across seven regions globally are obtained from Ref.~\cite{NASAFIRMS}, including data from Southeast Australia, Central Brazil, Eastern China, Northern Egypt, Central Germany, Western Turkey, and the south-central United States of America. For the sake of a diversity of wildfire data, all wildfires from August 1st, 2024, until the time of each passive observation within these seven regions are accounted for. As such, the mission horizon begins with $24$ active wildfires and ends with $81$ active wildfires. Additionally, only fires of a larger-than-average area, brightness, and distance from each grouping of fires, which are potentially repeat detections by VIIRS, as well as nominal or high confidence, are included in the dataset. As such, Fig.~\ref{fig:Wildfires} depicts each fire across all seven regions. 

\subsection{Results}

The results of the computational experiment are detailed through the performance of the WildFIRE-DS algorithm with various implementations in the Detection Module and Schedule Module. We present the following results for analysis: 1) the effectiveness of the Schedule Module in gathering data on the priority target set, 2) the accuracy of the Detection Module in combination with the Multi-Pass Confidence Module, 3) the performance of the EOSSP against the \REOSSP in the Schedule Module, and 4) the performance of the Late Fusion Model against the Early Fusion Model in the Detection Module. The next subsection provides a final discussion regarding the direct differences in CNN model implementation within the Detection Module and the scheduler used in the Schedule Module, additionally giving commentary on which combination is the most effective. 

The results of the WildFIRE-DS while using both the EOSSP and \REOSSP in the Schedule Module for each Block, when using the Early Fusion Model, are depicted in Table~\ref{tab:schedule_early}, reporting that the \REOSSP gathered \SI{220.25}{\%} more data than the EOSSP. The increase in the amount of data gathered is a result of both the difference in the priority target set of detected wildfires and the flexible nature of constellation reconfigurability in the \REOSSP. Additionally, since the priority target set of each schedule is only growing throughout the duration of the mission horizon, it is logical that both schedules gather an increasing amount of data in each subsequent Block, as more targets are considered for observation. Similarly, as the size of the priority target set grows and the number of observations and downlinks in each Block increases, the amount of battery power required to perform such operations increases. Furthermore, since the \REOSSP is gathering more data and thus performing more operations, the battery power required to operate the \REOSSP schedule is proportionally greater than that of the EOSSP schedule. Finally, as a partial maneuver budget is provided to Blocks $2$ and $3$ while the remaining full maneuver budget is provided to Block $4$, the total \REOSSP schedule uses \SI{1.71}{km/s} (\SI{85.48}{\%} of the total \SI{2.00}{km/s} provided). 

\begin{table}[!ht]
    \centering
    \caption{Schedule results of each Block (Early Fusion Model).}
    \resizebox{\textwidth}{!}{ 
    \begin{tabular}{ l l r r r r r }
\hline \hline
        Schedule & Block & $z$ & Data gathered, GB & Battery used, kJ & Provided maneuver budget, m/s & Maneuver costs, m/s \\
\hline
\REOSSP & 1 & 0.00    & 0.00  & 6052.00    & 1000.00 & 0.00    \\
        & 2 & 268.41  & 0.90  & 6241.66    & 1000.00 & 918.30  \\
        & 3 & 472.18  & 7.90  & 7382.56    & 540.85  & 436.08  \\
        & 4 & 796.86  & 16.50 & 8932.90    & 645.61  & 355.24  \\
Sum     &   & 1537.46 & 25.30 & \num{28609.12} &         & 1709.63 \\
\hline
EOSSP   & 1 & 0.00    & 0.00  & 6048.00    &         &         \\
        & 2 & 25.00   & 0.80  & 6203.94    &         &         \\
        & 3 & 56.00   & 1.90  & 6363.48    &         &         \\
        & 4 & 156.00  & 5.20  & 6955.92    &         &         \\
Sum     &   & 237.00  & 7.90  & \num{25571.34} &         &         \\
\hline \hline
    \end{tabular}
    }
    \label{tab:schedule_early}
\end{table} 

\begin{figure}[!ht]
    \centering
    \includegraphics[width = \textwidth]{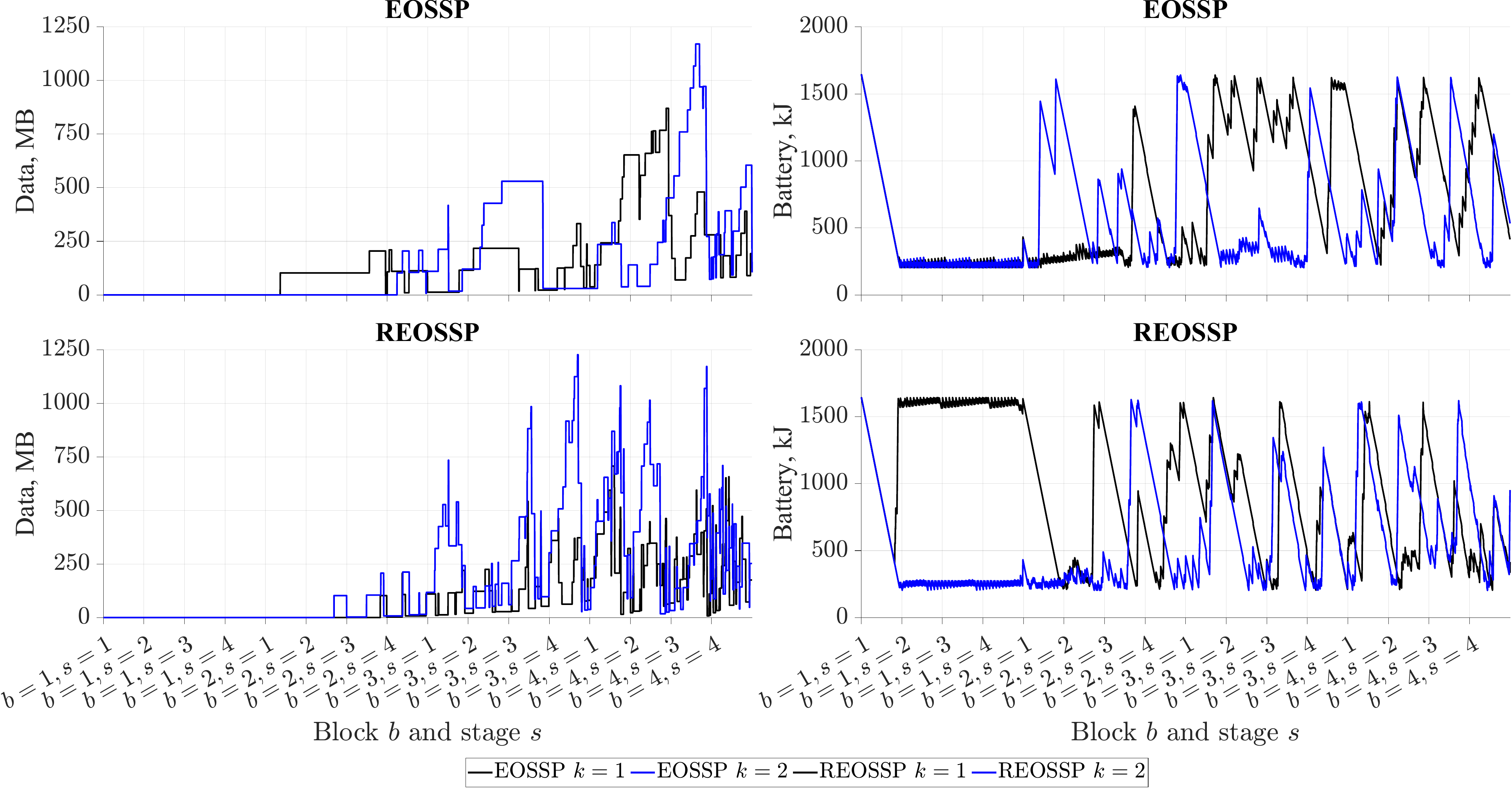}
    \caption{Progression of data and battery for the \textcolor{myblue}{\textsf{REOSSP}} and EOSSP (Early Fusion Model).}
    \label{fig:progression_early}
\end{figure}

The progression of the data and battery storage of the \REOSSP and EOSSP satellites is depicted in Fig.~\ref{fig:progression_early}, where the left portion shows the observation and downlink of data through the amount of occupied data storage for each satellite, and the right portion shows the usage and charging of the onboard battery for each satellite. The left portion of Fig.~\ref{fig:progression_early} further demonstrates that the satellites of the \REOSSP are capable of gathering more data than the satellites of the EOSSP, where Satellite $2$ of the \REOSSP peaks at \SI{1227.50}{MB} while Satellite $2$ of the EOSSP peaks at only \SI{1170.00}{MB}. Furthermore, the total number of observations of the \REOSSP and EOSSP satellites is $252$ and $79$, respectively, directly a result of the flexibility provided to the \REOSSP via orbital maneuvering. Furthermore, the rapid observation enabled by the flexibility provided to the \REOSSP allows for high-confidence auxiliary targets to be revisited sooner, and thus added to the priority target set for observation at an earlier time. Additionally, the right portion of Fig.~\ref{fig:progression_early} further demonstrates that the satellites of the \REOSSP require more battery power than the satellites of the EOSSP through the larger number of charging occurrences that take place, while the large peaks in both battery plots are simply a result of the large number of feasible charging schedules. 

\begin{figure}[!ht]
    \centering
    \includegraphics[width = \textwidth]{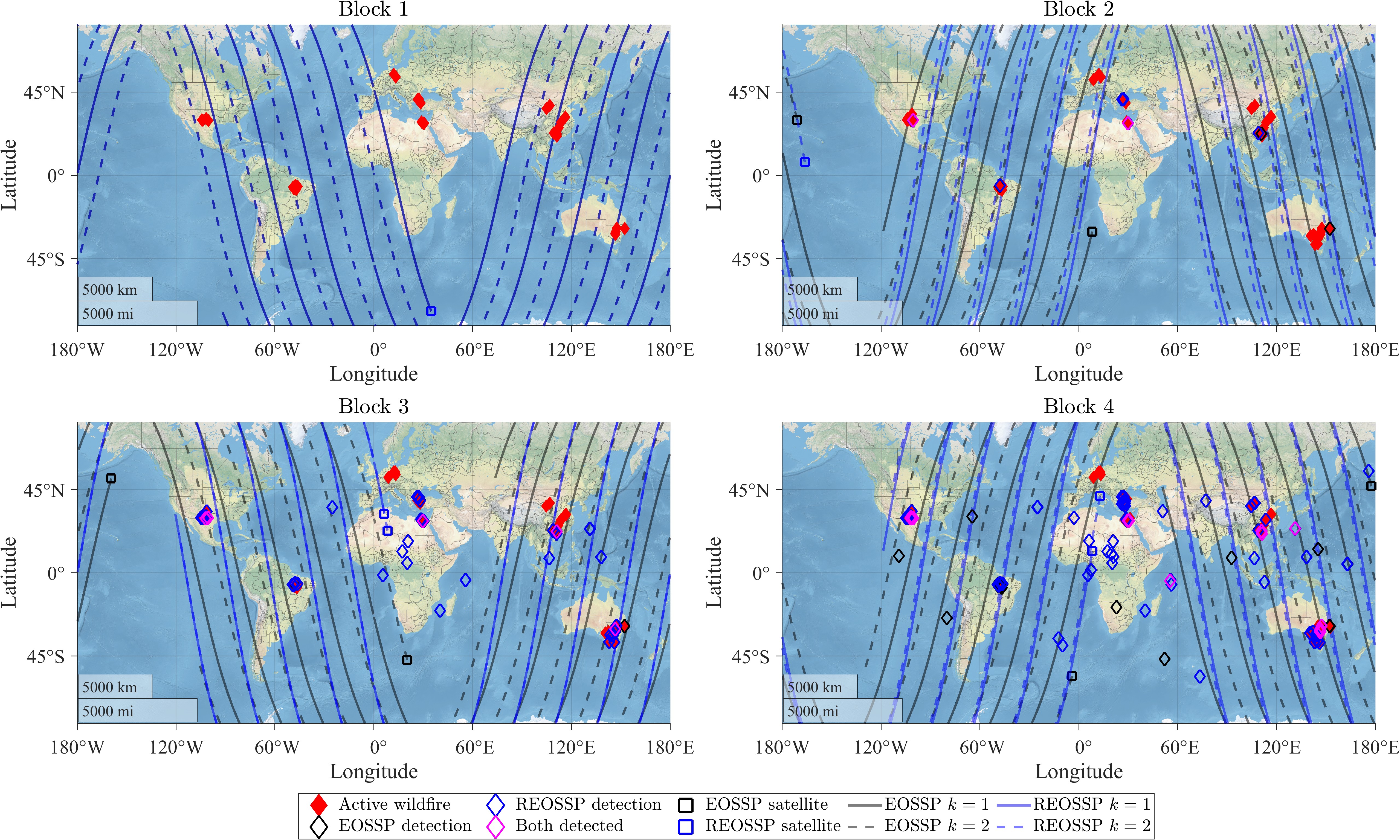}
    \caption{Updated active wildfire detections and satellite ground track in each Block (Early Fusion Model).}
    \label{fig:PriorityTargetSet_early}
\end{figure}

The updated priority target set of each scheduling problem and the satellite ground tracks of each Block are shown in Fig.~\ref{fig:PriorityTargetSet_early}. Within Fig.~\ref{fig:PriorityTargetSet_early}, red diamonds represent active wildfires, blue hollow diamonds represent the priority target set of the \REOSSP, black hollow diamonds represent the priority target set of the EOSSP, and magenta hollow diamonds represent mutual Detection Module interpretations via the Early Fusion Model (targets shared in the priority target set of both schedules). Additionally, the satellites of the \REOSSP and EOSSP are blue and black hollow squares, respectively, while the respective orbits of each satellite are represented as similarly colored lines. Each depiction refers to the priority target set at the end of the respective Block, thus being after all Detection Module interpretations and Multi-Pass Confidence Module updates, while priority target set detections throughout the entire Block are represented. At the start of Block $1$, the priority target set of each scheduling problem is empty, so no orbital maneuvers occur and the ground tracks of the EOSSP and \REOSSP are identical. Table~\ref{tab:Detections_early} reflects the number of active wildfires detectable within a Block, as well as the results of the Schedule Module execution of the EOSSP and \REOSSP per Block relative to the number of true-positive wildfire detections. Table~\ref{tab:Detections_early} additionally reflects that the priority target set is not expanded upon within Block $1$, a result of the high priority target threshold of \SI{95}{\%}. 

\begin{table}[!ht]
    \centering
    \caption{Detection status of each Block (Early Fusion Model).}
    \resizebox{\textwidth}{!}{ 
    \begin{tabular}{ l l r r r r r r }
\hline \hline
        &  & \multicolumn{3}{c}{EOSSP} & \multicolumn{3}{c}{\REOSSP} \\
        \cmidrule(lr){3-5} \cmidrule(lr){6-8}
        Block & Cumulative       & Detections & True positives & Useful data/  & Detections & True positives & Useful data/ \\
              & active wildfires &            &                & Obtained data, GB &            &                & Obtained data, GB \\
\hline
Block 1 & 36 & 0  & ~ & ~ & 0 & ~ & ~ \\
Block 2 & 58 & 5  & 5, \SI{100.00}{\%}  & 0.80/\SI{0.80}{} & 9   & 9, \SI{100.00}{\%}  & 0.90/\SI{0.90}{}   \\
Block 3 & 69 & 14 & 14, \SI{100.00}{\%} & 1.90/\SI{1.90}{} & 57  & 44, \SI{77.19}{\%}  & 6.10/\SI{7.90}{}   \\
Block 4 & 81 & 46 & 35, \SI{76.09}{\%}  & 3.95/\SI{5.20}{} & 160 & 123, \SI{76.88}{\%} & 12.68/\SI{16.50}{} \\
\hline
Sum     &    &    &                     & 6.65/\SI{7.90}{} &     &                     & 19.68/\SI{25.30}{} \\
\hline \hline
    \end{tabular}
    }
    \label{tab:Detections_early}
\end{table}

As the mission progresses, the \REOSSP is provided with more knowledge regarding the most confident interpretations within the auxiliary target set, allowing for more rapid revisits and subsequent re-interpretations of the same geolocations. As such, the \REOSSP acquires more priority targets than the EOSSP, as well as at an earlier time in the mission duration, but additionally, the \REOSSP acquires a larger number of false-positive detections within the priority target set. Although the ratio of true-positive detections in the \REOSSP is still higher than the EOSSP, a fact largely resulting from the Multi-Pass Confidence Module filtering many false-positive detections from entering the priority target set. Examples of false-positive detections that present themselves within the priority target set are shown in Appendix~B. Additionally, the number of true-positive detections obtained by the \REOSSP is higher than the number of active wildfires in Block $4$ due to rounding errors in the geometric determination of wildfire locations in satellite imagery, shown in Appendix~A. Due to said rounding errors, the same wildfire may be detected in multiple images with small differences in pixel location, satellite position, or a combination of both, leading to a slightly different geographic position for the detected wildfire of approximately half a degree in latitude or longitude. These repeat detections of the same wildfire are still treated as true-positive detections, as the difference in the detected location is small enough that the actual wildfire location would likely still fall within the satellite FOV, and the \REOSSP must choose a single target within a cluster of repeat detections for observation. 

The \REOSSP satellites select widely different orbital slot options in Block $2$, as shown by the lack of overlap in the ground tracks displayed in Fig.~\ref{fig:PriorityTargetSet_early}, while the orbital slot options are extremely similar in Block $3$ and $4$, wherein the satellite ground tracks overlap heavily, and the satellites themselves are closely trailing one another, particularly in Block $3$. The EOSSP satellite ground tracks are consistent between each Block, providing a consistent yet often slow revisit time relative to the priority target set, thus leading to a lower amount of gathered data depicted in Table~\ref{tab:schedule_early}, and relative to the auxiliary target set, thus leading to a lower number of targets being added to the priority target set. The differences in the orbital slot option of the \REOSSP satellites result in better placement of each satellite such that visibility of the auxiliary target set and priority target set is provided more frequently. 

\begin{figure}[!ht]
    \centering
    \includegraphics[width = \textwidth]{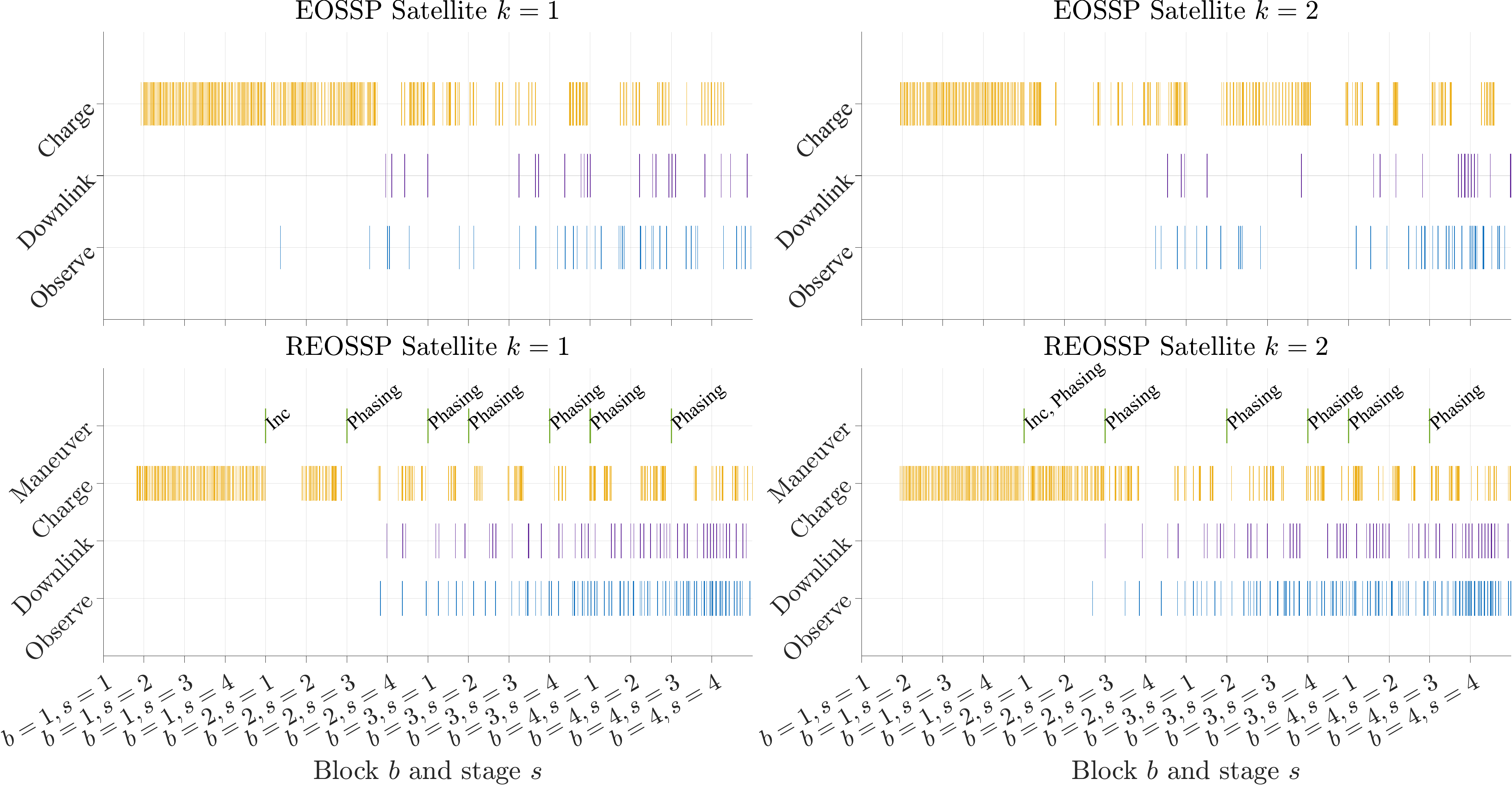}
    \caption{Task execution timing for the \textcolor{myblue}{\textsf{REOSSP}} and EOSSP (Early Fusion Model).}
    \label{fig:Execution_early}
\end{figure}

To further exemplify the results depicted in Figs.~\ref{fig:progression_early} and~\ref{fig:PriorityTargetSet_early}, as well as Tables~\ref{tab:schedule_early} and~\ref{tab:Detections_early}, Fig.~\ref{fig:Execution_early} shows the execution of tasks by the satellites in the \REOSSP and EOSSP throughout the mission duration. The top half of Fig.~\ref{fig:Execution_early} reports the task execution of the EOSSP satellites, including solar charging, data downlink, and priority target observation, while the bottom half of Fig.~\ref{fig:Execution_early} reports the task execution of the \REOSSP satellites, additionally including orbital maneuvering alongside the type of maneuver performed when a new orbital slot option is selected. As for the maneuvers performed by each satellite in the \REOSSP, both satellites perform inclination lowering maneuvers at the beginning of Block $2$, lowering from \SI{98.80}{deg} to \SI{95.91}{deg} and from \SI{98.79}{deg} to \SI{95.90}{deg} concerning Satellite $1$ and Satellite $2$, respectively. Additionally, each satellite performs occasional phasing maneuvers for the remainder of the mission duration, ensuring that the satellites remain in the best possible position for target observation. From these results, Fig.~\ref{fig:Execution_early} provides further evidence that the number and density of observations and data downlink occurrences are higher in the \REOSSP than in the EOSSP. As such, the effectiveness in identifying priority targets provided by the additional flexibility in the \REOSSP and the confidence increase provided by the Multi-Pass Confidence Module leads to more accurate and prompt wildfire identification, as well as the ability to obtain more data regarding the identified wildfires. 

Now pivoting to the results of the WildFIRE-DS when using the Late Fusion Model, as shown in Table~\ref{tab:schedule_late}; the results indicate that the \REOSSP gathered \SI{118.42}{\%} more data than the EOSSP. The improvement provided by the \REOSSP with the Late Fusion Model is \SI{46.23}{\%} lower than that of the Early Fusion Model, and the amount of data itself is \SI{67.19}{\%} lower. However, battery usage still increases proportionally with the number of required operations within the \REOSSP. Additionally, the total \REOSSP schedule uses \SI{1.92}{km/s} (\SI{95.99}{\%} of the total \SI{2.00}{km/s} provided), a \SI{12.29}{\%} maneuver cost increase over the \REOSSP schedule using the Early Fusion Model in the Detection Module. 

\begin{table}[!ht]
    \centering
    \caption{Schedule results of each Block (Late Fusion Model).}
    \resizebox{\textwidth}{!}{ 
    \begin{tabular}{ l l r r r r r }
\hline \hline
        Schedule & Block & $z$ & Data gathered, GB & Battery used, kJ & Provided maneuver budget, m/s & Maneuver costs, m/s \\
\hline
\REOSSP & 1 & 0.00   & 0.00 & 6052.00  & 1000.00 & 0.00    \\
        & 2 & 207.08 & 0.20 & 6086.92  & 1000.00 & 884.44  \\
        & 3 & 218.42 & 1.70 & 6348.82  & 557.78  & 521.31  \\
        & 4 & 374.39 & 6.40 & 7169.44  & 594.25  & 513.96  \\
Sum     &   & 799.89 & 8.30 & \num{25657.18} &         & 1919.71 \\
\hline
EOSSP   & 1 & 0.00   & 0.00 & 6048.00  &         &         \\
        & 2 & 10.00  & 0.30 & 6116.64  &         &         \\
        & 3 & 32.00  & 1.10 & 6223.80  &         &         \\
        & 4 & 73.00  & 2.40 & 6483.30  &         &         \\
Sum     &   & 115.00 & 3.80 & \num{24871.74} &         &      \\
\hline \hline
    \end{tabular}
    }
    \label{tab:schedule_late}
\end{table} 

\begin{figure}[!ht]
    \centering
    \includegraphics[width = \textwidth]{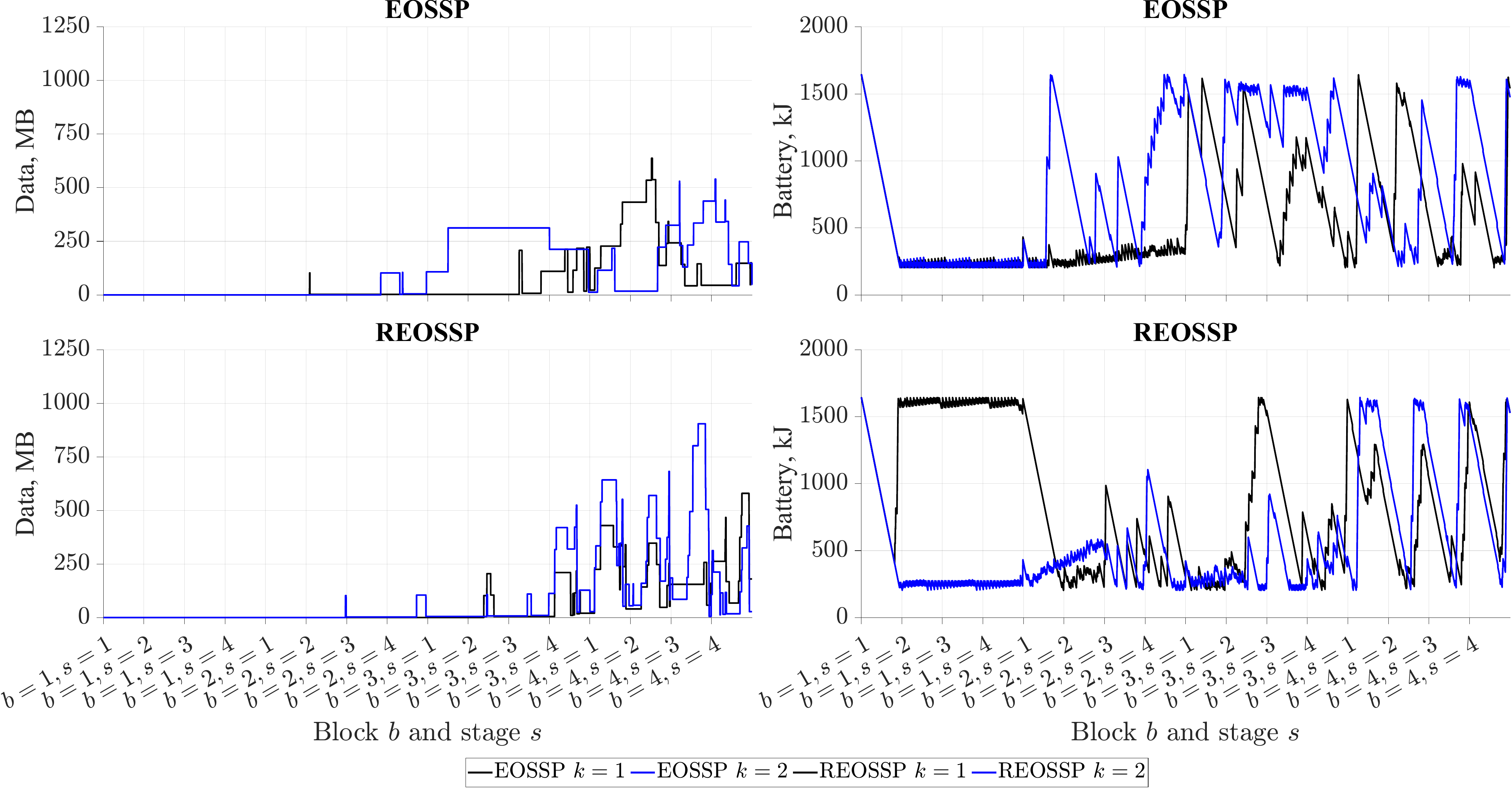}
    \caption{Progression of data and battery for the \textcolor{myblue}{\textsf{REOSSP}} and EOSSP (Late Fusion Model).}
    \label{fig:progression_late}
\end{figure}

Additionally, the progression of the data and battery storage of the EOSSP and \REOSSP satellites is depicted in Fig.~\ref{fig:progression_late}. Again, the \REOSSP satellites are capable of gathering more data than the satellites of the EOSSP, where Satellite $2$ of the \REOSSP peaks at \SI{905}{MB} while Satellite $1$ of the EOSSP peak at \SI{637.50}{MB}, a decrease of \SI{26.27}{\%} and \SI{45.51}{\%} compared to the use of the Early Fusion Model, respectively. Similarly, the total number of observations of the \REOSSP and EOSSP satellites is $83$ and $39$, respectively, a proportionate decrease with the lower amount of data obtained. As previously stated, the large peaks and variability in the right portion of Fig.~\ref{fig:progression_late} are a result of the large number of feasible charging schedules. 

\begin{figure}[!ht]
    \centering
    \includegraphics[width = \textwidth]{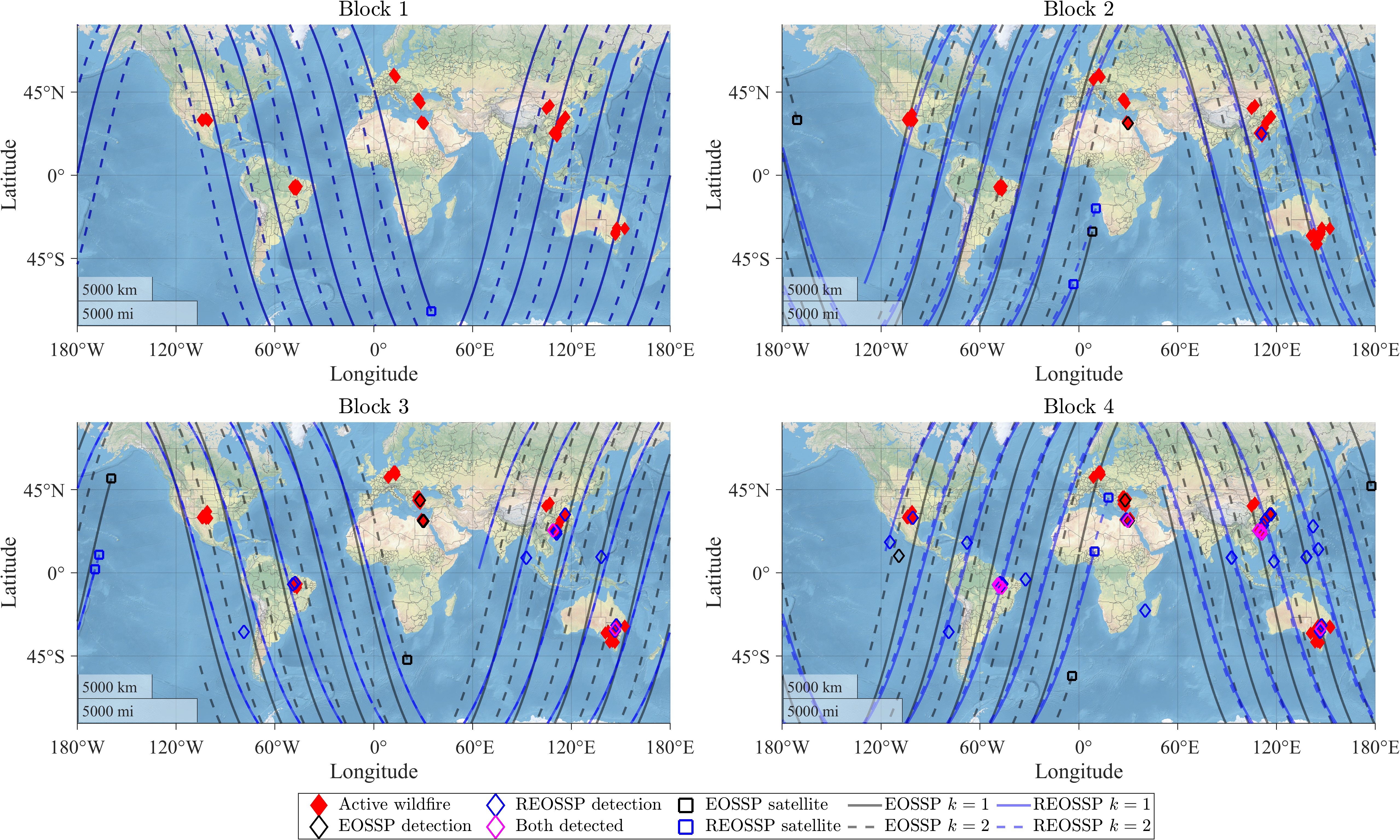}
    \caption{Updated active wildfire detections and satellite ground track in each Block (Late Fusion Model).}
    \label{fig:PriorityTargetSet_late}
\end{figure}

Separately, Fig.~\ref{fig:PriorityTargetSet_late} shows the updated priority target set of each scheduling problem alongside satellite ground tracks, with the inclusion of orbital maneuvers within the \REOSSP. Subsequently, Table~\ref{tab:Detections_late} reports the results of the Schedule Model execution relative to the number of true-positive wildfire detections within the priority target set. Similar to the use of the Early Fusion Model, the use of the Late Fusion Model results in the priority target set not being expanded upon in Block $1$. As a result of the flexibility provided by the \REOSSP, more priority targets are acquired at an earlier time than with the EOSSP, although with the use of the Late Fusion Model, a lower ratio of true-positive and overall detections is provided. Contrarily, the EOSSP has a \SI{100}{\%} true-positive detection rate in Blocks $2$ and $3$, with a \SI{95}{\%} true-positive detection rate in Block $4$, while only incurring less than half of the overall detections when compared to the use of the Early Fusion Model, thus still acquiring \SI{52.55}{\%} less useful data. As such, the Multi-Pass Confidence Module is still operating effectively to drive mixed and consistently low-confidence interpretations to zero; however, since the Late Fusion Model has lower confidence values than the Early Fusion Model on average, the number of targets in the priority target set is proportionally lower. Similar to the orbital slot option selection when using the Early Fusion Model, the \REOSSP satellites select similar orbital slot options in all Blocks after Block $1$, as shown by how closely each ground track overlaps in Fig.~\ref{fig:PriorityTargetSet_early} and how closely the satellites are trailing one another. 

\begin{table}[!ht]
    \centering
    \caption{Detection status of each Block (Late Fusion Model).}
    \resizebox{\textwidth}{!}{ 
    \begin{tabular}{ l l r r r r r r }
\hline \hline
        &  & \multicolumn{3}{c}{EOSSP} & \multicolumn{3}{c}{\REOSSP} \\
        \cmidrule(lr){3-5} \cmidrule(lr){6-8}
        Block & Cumulative       & Detections & True positives & Useful data/  & Detections & True positives & Useful data/ \\
              & active wildfires &            &                & Obtained data, GB &            &                & Obtained data, GB \\
\hline
Block 1 & 36 & 0  & ~ & ~ & 0 & ~ & ~ \\
Block 2 & 58 & 1  & 1, \SI{100.00}{\%} & 0.30/\SI{0.30}{} & 1  & 1, \SI{100.00}{\%} & 0.20/\SI{0.20}{} \\
Block 3 & 69 & 5  & 5, \SI{100.00}{\%} & 1.10/\SI{1.10}{} & 13 & 10, \SI{76.92}{\%} & 1.31/\SI{1.70}{} \\
Block 4 & 81 & 20 & 19, \SI{95.00}{\%} & 2.28/\SI{2.40}{} & 39 & 27, \SI{69.23}{\%} & 4.43/\SI{6.40}{} \\ 
\hline
Sum     &    &    &                    & 3.68/\SI{3.80}{} &    &                    & 5.94/\SI{8.30}{} \\
\hline \hline
    \end{tabular}
    }
    \label{tab:Detections_late}
\end{table}

\begin{figure}[!ht]
    \centering
    \includegraphics[width = \textwidth]{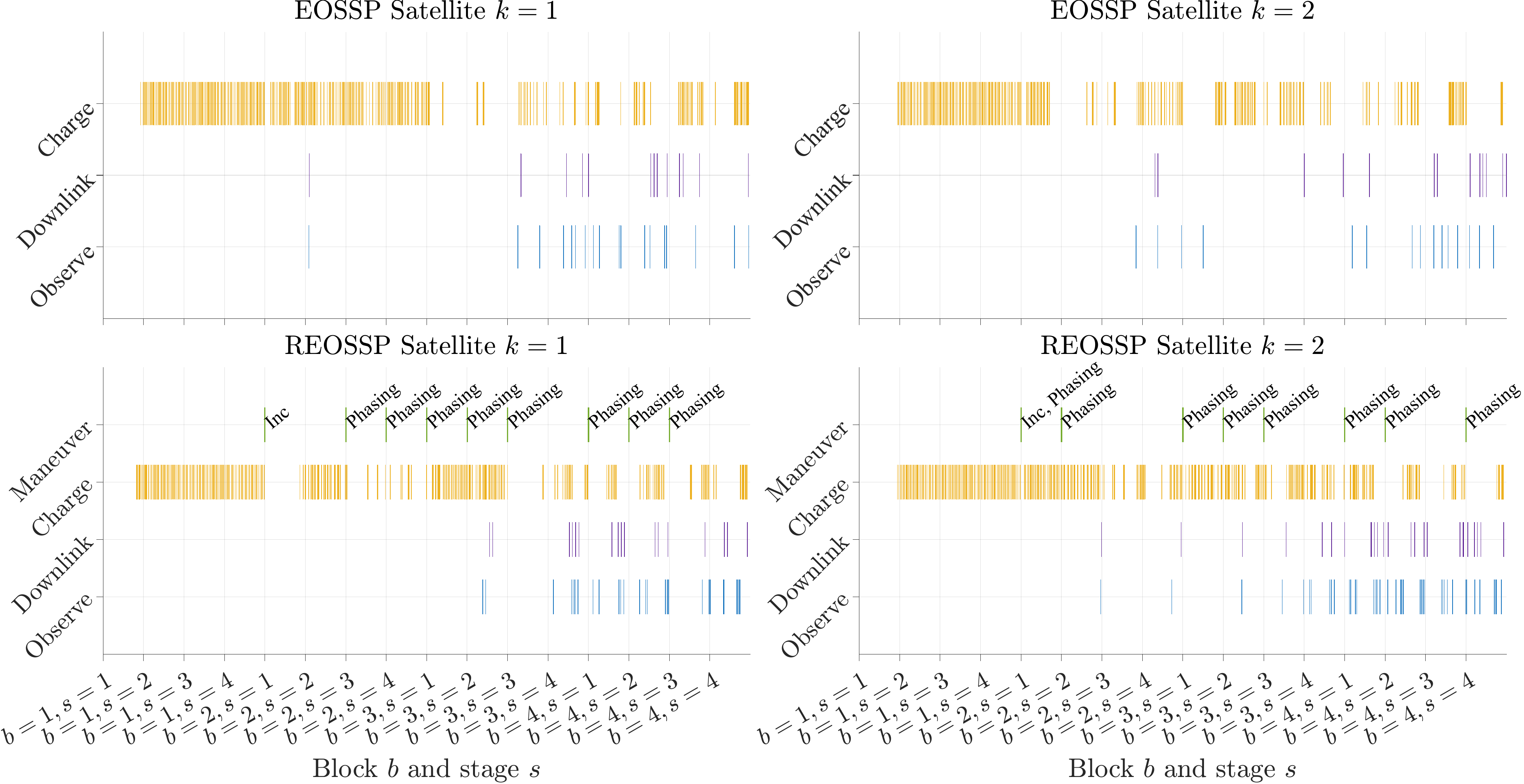}
    \caption{Task execution timing for the \textcolor{myblue}{\textsf{REOSSP}} and EOSSP (Late Fusion Model).}
    \label{fig:Execution_late}
\end{figure}

Finally, to emphasize the results depicted in Figs.~\ref{fig:progression_late} and~\ref{fig:PriorityTargetSet_late}, as well as Tables~\ref{tab:schedule_late} and~\ref{tab:Detections_late}, Fig.~\ref{fig:Execution_late} shows the execution of tasks by the satellites in the \REOSSP and EOSSP in the same manner as Fig.~\ref{fig:Execution_late}. Similar to the use of the Early fusion model, each satellite begins Block $2$ with a change in inclination, although in this case each satellite raises in inclination from \SI{98.80}{deg} to \SI{101.69}{deg} and \SI{98.79}{deg} to \SI{101.68}{deg} for Satellite $1$ and Satellite $2$, respectively. As a result of optimal maneuvering, the \REOSSP satellites perform more priority target observations and data downlinks than the EOSSP satellites, despite Satellite $1$ of the EOSSP performing an observation long before Satellite $1$ of the \REOSSP. 

\subsection{Discussions}

With the presentation of the WildFIRE-DS results regarding the Early Fusion Model and Late Fusion Model within the Detection Module, as well as the comparison between the EOSSP and \REOSSP in the Schedule Module throughout, some conclusions can be drawn regarding which combination in each Module is most effective. Table~\ref{tab:Full_Results} reports what the authors have determined to be the most important metrics of the WildFIRE-DS results relative to each combination between the Detection Module and Scheduling Module. These three metrics include: 1) the number of true-positive wildfire detections, reflecting the effectiveness of the Detection Module and Multi-Pass Confidence Module, and the responsiveness of the Scheduling Module, 2) the total amount of obtained data in GB, reflecting the performance of the Scheduling Module relative to the number of overall detections via the Detection Module, and 3) the maneuver costs of the \REOSSP schedule, displaying the overall cost incurred to obtain the results of the Schedule Module. 

The results reported in Table~\ref{tab:Full_Results} are in line with expectations from the CNN model comparative analysis in Sec.~\ref{subsec:CNN} and with previous research on constellation reconfigurability \cite{Pearl2024Benchmarking,Pearl2025REOSSP,Pearl2025Developing}. First, it was hypothesized that the Early Fusion Model is the most effective CNN model, outperforming the Late Fusion Model. Indeed, Table~\ref{tab:Full_Results} reports that using the Early Fusion Model compared to the Late Fusion Model increases the data obtained by \SI{107.89}{\%} for the EOSSP and by \SI{204.82}{\%} for the \REOSSP. Second, it has been shown that the \REOSSP outperforms the EOSSP \cite{Pearl2025Developing,Pearl2025REOSSP}. Similarly, Table~\ref{tab:Full_Results} reports that using the \REOSSP compared to the EOSSP increases the data obtained by \SI{220.25}{\%} for the Early Fusion Model and by \SI{118.42}{\%} for the Late Fusion Model. As such, the combination of the Early Fusion Model within the Detection Module and the \REOSSP within the Schedule Module is the most effective iteration of the WildFIRE-DS, achieving both the highest amount of data and number of true-positive wildfire detections, while incurring the lowest maneuver costs. Additionally, the combination of the Late Fusion Model within the Detection Model and the EOSSP within the Schedule Module is the least effective iteration of the WildFIRE-DS, while incurring no maneuver costs in the EOSSP, the number of true-positives and amount of obtained data is the lowest of all WildFIRE-DS iterations.

\begin{table}[!ht]
    \centering
    \caption{Important metrics for comparison within the WildFIRE-DS algorithm.}
    \resizebox{\textwidth}{!}{ 
    \begin{threeparttable}
        \begin{tabular}{ l c c c c c }
\hline \hline 
Detection Module   & \multicolumn{5}{c}{Schedule Module}\\
\cmidrule(lr){2-6}
& \multicolumn{2}{c}{EOSSP} & \multicolumn{3}{c}{\REOSSP} \\
\cmidrule(lr){2-3} \cmidrule(lr){4-6}
& True positives & Obtained data, GB & True positives & Obtained data, GB & Maneuver costs, m/s \\
\hline 
Early Fusion Model & 35 & 7.90 & \textbf{123}\tnote{a} & \textbf{25.30} & \textbf{1709.63} \\
\hline
Late Fusion Model  & \textit{19}\tnote{b} & \textit{3.80} & 27 & 8.30  & 1919.71 \\
\hline \hline
        \end{tabular}
        \begin{tablenotes}
           \item[a] A \textbf{bold} entry denotes the best results from the module combinations.
           \item[b] An \textit{italic} entry denotes the worst results from the module combinations.
        \end{tablenotes}
    \end{threeparttable}
    }
    \label{tab:Full_Results}
\end{table}

Overall, the WildFIRE-DS algorithm automates the pipeline of wildfire detection and subsequent satellite scheduling with the inclusion of recent improvements to both aspects of the pipeline as a proof of concept of the proposed framework. The ability of the WildFIRE-DS algorithm to autonomously detect wildfires provided by sensor fusion techniques and CNN models in the Detection Module, perform statistical updates from subsequent CNN model interpretations in the Multi-Pass Confidence Module, and responsively schedule a constellation of satellites in the Schedule Module enables fully autonomous operations. Such autonomy increases the pace of the decision-making process both through the removal of a human decision-maker and through improvements to the detection and scheduling aspects of the overall pipeline, but may still allow an operator to intervene when desired by an end user. Additionally, the modularity of the WildFIRE-DS is shown through the highly flexible nature of the \REOSSP in the Schedule Module provided through the implementation of constellation reconfigurability and the effective high-confidence interpretations provided by the Early Fusion Model in the Detection Module. Moreover, the inclusion of the Multi-Pass Confidence Module in tandem with the Detection Module allows for the filtering of low-confidence Detection Module interpretations, ensuring that only consistently high-confidence detections are considered for the priority target set and subsequent satellite task scheduling in the Schedule Module. The autonomous operation of the WildFIRE-DS, with inherent modularity, is shown through the progression of the priority target set over the mission horizon, the increase in and rapid obtainment of gathered data by the \REOSSP over the baseline EOSSP, the scheduled task execution of each scheduling in the Schedule Module, and the effectiveness of the Multi-Pass Confidence Module in filtering false-positive detections from the Detection Module.

\section{Conclusion} \label{sec:Conclusion}

This paper proposes a cohesive framework to automate the detection of active wildfires and the subsequent scheduling of a constellation of maneuverable satellites, alongside developing the WildFIRE-DS algorithm as a proof of concept. The WildFIRE-DS realizes the proposed framework through the use of three key Modules: employing the YOLOv11 CNN architecture in the Early Fusion Model and Late Fusion Model within the Detection Module, Bayesian statistics to update interpretation confidence in the Multi-pass Detection Module, and the EOSSP and \REOSSP from Refs.~\cite{Pearl2025Developing,Pearl2025REOSSP} in the Schedule Module. The Early Fusion Model and Late Fusion Model are trained and validated using simulated satellite imagery gathered from STK \cite{STK}, to identify active wildfires and return the location of detections, while additionally employing sensor fusion to increase the effectiveness over the Band $6$ Model and Band $7$ Model. In tandem, the Multi-Pass Confidence Module leverages Bayesian statistics to incorporate the maximum amount of available information on each wildfire detection, thus allowing multiple flyovers to contribute new interpretation confidence data throughout the mission duration and successfully filtering a large number of false-positive detections. Separately, the Schedule Module makes use of the \REOSSP and EOSSP to consider available data and battery storage, as well as target observation, data downlink, solar charging, and orbital maneuvering to obtain provable optimal schedule solutions. The WildFIRE-DS allows autonomy over a given mission horizon through the division into a series of Blocks, during which a Schedule Module duration takes place, allowing passive observations to be interpreted by the Detection Module and allowing the Multi-Pass Confidence Module to update a priority target set of active wildfires, which is subsequently used by the Schedule Module for near-real-time schedule updates and scheduling of subsequent Blocks. 

The computational experiment, conducted in Sec.~\ref{sec:Experiment} using real-world wildfire information obtained from NASA FIRMS \cite{NASAFIRMS}, demonstrates the capability of the WildFIRE-DS algorithm to detect active wildfires and schedule optimal observations of the detected wildfires. In the computational experiment, the comparison between the various combinations of the Early Fusion Model, Late Fusion Model, EOSSP, and \REOSSP is provided, where the \REOSSP is shown to collect \SI{220.25}{\%} and \SI{118.42}{\%} more data than the EOSSP relative to the Early Fusion Model and Late Fusion Model, respectively. Furthermore, the use of the Early Fusion Model in the Detection Module consistently results in not only a higher number of overall detections but also a better ratio of true-positive wildfire detections overall, thus resulting in more useful data obtained throughout the mission. As such, the high level of performance obtained by the combination of the Early Fusion Model and \REOSSP is provided through the effectiveness of the Early Fusion Model in obtaining high-confidence image interpretation, which is further refined by the Multi-Pass Confidence Module, and the increased flexibility in the \REOSSP provided by the constellation reconfigurability with multiple benefits. The benefits of the flexibility of constellation reconfigurability include the provision of more early and frequent opportunities for priority target set observation, such that high-confidence interpretations are added to the priority target set for scheduling sooner in the mission horizon. As evidence of the flexibility of the satellites in the \REOSSP, each performs separate types of orbital maneuvers, including lowering inclination, raising RAAN, and performing phasing at each maneuver. Overall, the WildFIRE-DS implements each module, the Early Fusion Model and Late Fusion Model, and the \REOSSP and EOSSP to operate in a mission horizon autonomously, allowing the obtainment of an otherwise infeasible amount of data regarding active wildfires.

Future endeavors may be undertaken to iterate on, modify the application of, and further investigate the WildFIRE-DS. Firstly, testing of the WildFIRE-DS and associated modules on real-world satellite hardware would provide the means to improve upon the computational load of the algorithm for the purpose of onboard implementation. Secondly, alternative scheduling methods for the \REOSSP and EOSSP may reduce the computation time currently required to obtain an optimal schedule within the Schedule Module, thus further improving the fidelity to implement the WildFIRE-DS onboard real-world hardware. Similarly, dedicated testing of the WildFIRE-DS onboard real-world hardware would allow verification of the algorithm and relative operating time scales, which could then be improved. A specific test that may prove particularly fruitful is the testing of the CNN models depicted in Sec.~\ref{subsec:CNN} on imagery produced by the imaging system developed in Ref.~\cite{Fukuhara2017Detection} due to the similarity to the images generated in STK. Separately, while this paper considers active wildfires, the Detection Module may be reoriented for the purpose of monitoring other natural disasters, such as tropical cyclones or active flooding events. Additionally, due to the nature of constellation reconfigurability assigning orbital maneuvers relative to known events, the optimal configuration may leave detection of further unknown events somewhat lacking. As such, performing an analysis of the opportunity cost of reconfiguring the constellation would provide further insight into the best maneuvers to perform for target observation and continued global monitoring simultaneously. Finally, the incorporation of either low-thrust orbital maneuvers \cite{McGrath2019General} to decrease the cost of reconfiguration operations or refueling operations \cite{Jonchay2022Refueling} as an aspect of the \REOSSP would increase the longevity of a reconfigurable constellation to provide operations over a potentially much longer time horizon.

\section*{Appendix A: Geometry to Determine Detected Wildfire Geolocations}

\begin{figure}[!ht]
    \centering
    \includegraphics[width = 0.5\textwidth]{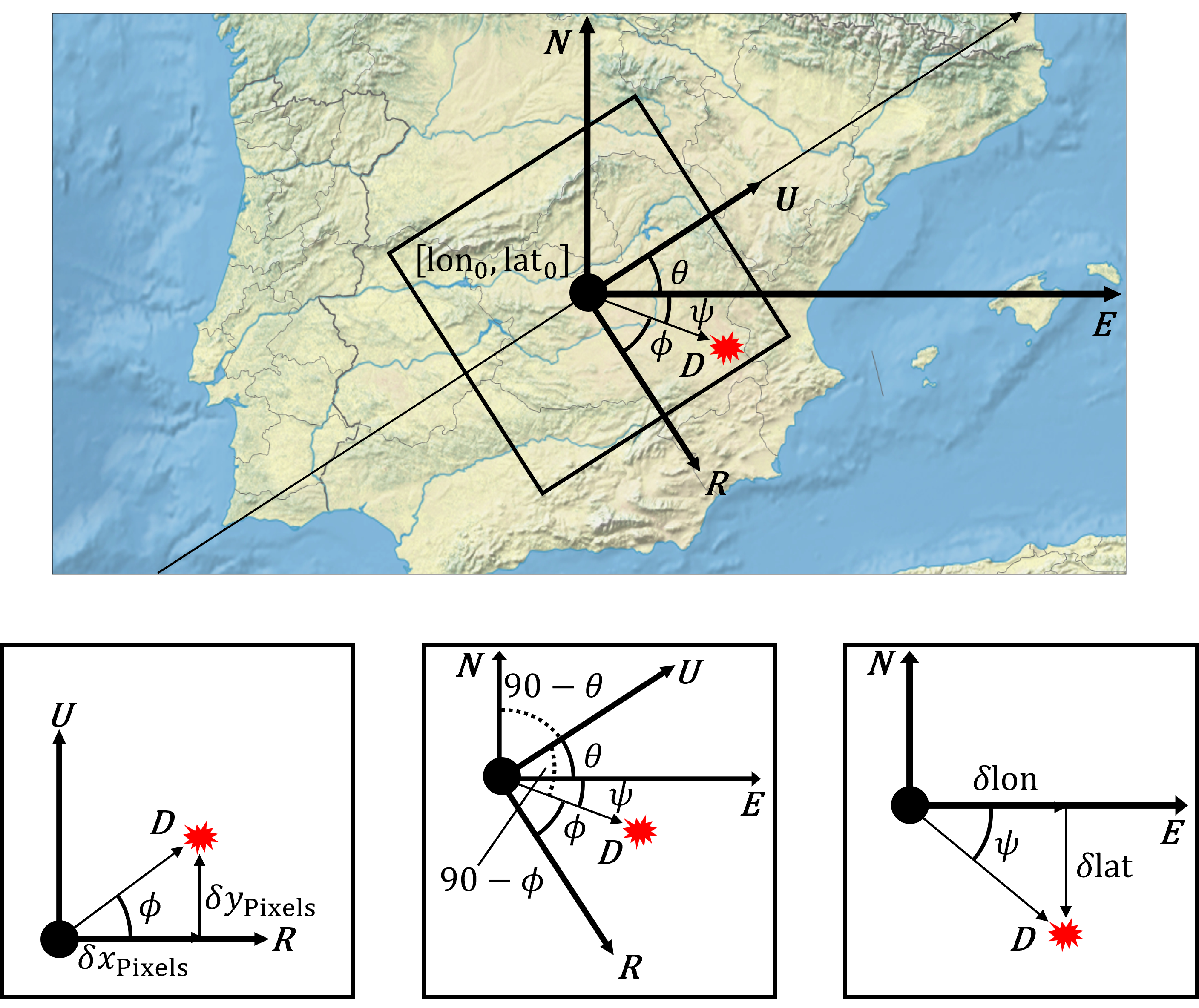}
    \caption{Geometric representation of detected wildfires.}
    \label{fig:Detection_Geometery}
\end{figure}

To compute the accurate geolocation (latitude and longitude) of detected wildfires within a satellite image, key geometric information is required. Initially, only the current satellite orbital parameters and the detected wildfire bounding-box location within the satellite image are known at any given time. Therefore, the following information is defined from known information: $\text{lat}_0$ and $\text{lon}_0$ are the initial latitude and longitude of the satellite (obtained from the satellite orbital parameters), respectively, $\bm{U}$ and $\bm{R}$ are the vertical and horizontal directions within the satellite image, respectively, and $\bm{D}$ is the directional vector from the center of the satellite image to the detected wildfire bounding box. With the provided information, $\delta x_{\text{Pixels}}$ and $\delta y_{\text{Pixels}}$ are the pixel distances from the center of the satellite image to $\bm{D}$ along $\bm{R}$ and $\bm{U}$ axes, respectively, with the angle $\phi$ from $\bm{R}$ to $\bm{D}$, as shown in the bottom left of Fig.~\ref{fig:Detection_Geometery}; the angle $\phi$ is computed in Eq.~\eqref{Geometry:phi}. Furthermore, $\theta$ is the angle from the Eastern direction to the satellite's inclined direction of travel (obtained from the satellite orbital parameters) as shown in the top of Fig.~\ref{fig:Detection_Geometery}, such that $\theta$ is the satellite inclination on the ascending half of the satellite orbit and the negative of the satellite inclination on the descending half of the satellite orbit, as computed in Eq.~\eqref{Geometry:theta}. Within Eq.~\eqref{Geometry:theta}, $v_z$ is the vertical component of the satellite velocity vector, which is used to indicate whether the satellite is in the ascending or descending half of the orbital path. Next, $\psi$ is the angle from the Eastern direction to $\bm{D}$, as shown in the bottom center of Fig.~\ref{fig:Detection_Geometery}, which is computed from $\theta$ and $\phi$ in Eq.~\eqref{Geometry:psi}. Finally, $\psi$ is used alongside $\text{lat}_0$ and $\text{lon}_0$ to compute $\delta \text{lat}$ and $\delta \text{lon}$, the difference in latitude and longitude, respectively, of the detected wildfire bounding box relative to the satellite geolocation, as shown in the bottom right of Fig.~\ref{fig:Detection_Geometery} and Eqs.~\eqref{Geometry:Delta_lat} and~\eqref{Geometry:Delta_lon}. This final set of equations involves the computation of the magnitude of $\bm{D}$ through the square root term $\sqrt{\delta x_{\text{Pixels}}^2 + \delta y_{\text{Pixels}}^2}$, the conversion from a distance in pixels to a distance in kilometers using the ground sample distance (GSD), and a conversion from kilometers to degrees latitude by a factor of $110.574$ and to degrees longitude by a factor of $111.320 \cos(\text{lat}_0 + \delta \text{lat})$. The GSD is computed for each satellite in STK \cite{STK}, dependent upon the satellite altitude, the size of the image in pixels, and the satellite sensor's size and effective focal length. 
\begin{subequations}
    \begin{align}
        \phi &= \arctan\left( \frac{\delta y_{\text{Pixels}}}{\delta x_{\text{Pixels}}} \right) 
        \label{Geometry:phi}\\
        \theta &= \begin{cases} i, & \quad \text{if } v_z \ge 0 \\ -i, & \quad \text{if } v_z < 0 \end{cases} 
        \label{Geometry:theta}\\
        \psi &= \theta + \phi - \SI{90}{\deg} 
        \label{Geometry:psi}\\
        \delta \text{lat} &= \frac{\text{GSD}\sqrt{\delta x_{\text{Pixels}}^2 + \delta y_{\text{Pixels}}^2} \sin(\psi)}{110.574} 
        \label{Geometry:Delta_lat} \\
        \delta \text{lon} &= \frac{\text{GSD}\sqrt{\delta x_{\text{Pixels}}^2 + \delta y_{\text{Pixels}}^2} \cos(\psi)}{111.320\cos(\text{lat}_0 + \delta \text{lat})}
        \label{Geometry:Delta_lon}
    \end{align}
    \label{Geometry}
\end{subequations}

\section*{Appendix B: Examples of False-Positive CNN Interpretations}

\begin{figure}[!ht]
    \centering
    \begin{subfigure}[h]{0.24\textwidth}
    \centering
        \includegraphics[width = \textwidth]{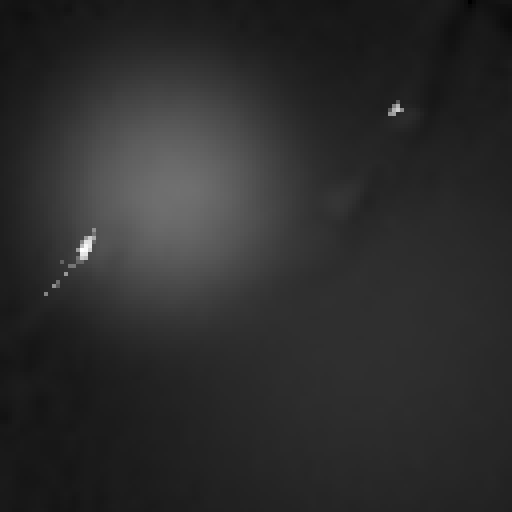}
        \caption{Band $6$ island image during daytime.}
        \label{fig:Band6_Island}
    \end{subfigure}
    \hfill
    \begin{subfigure}[h]{0.24\textwidth}
    \centering
        \includegraphics[width = \textwidth]{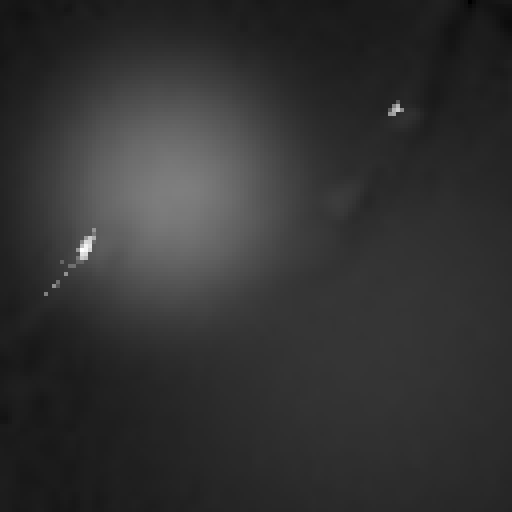}
        \caption{Band $7$ island image during daytime.}
        \label{fig:Band7_Island}
    \end{subfigure}
    \hfill
    \begin{subfigure}[h]{0.24\textwidth}
    \centering
        \includegraphics[width = \textwidth]{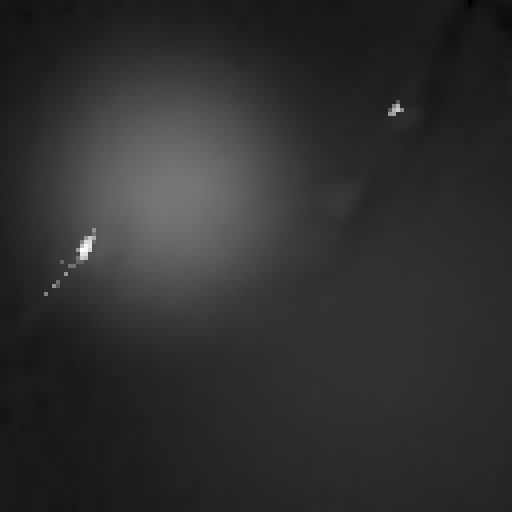}
        \caption{Early fusion island image during daytime.}
        \label{fig:Early_Island}
    \end{subfigure}
    \hfill
    \begin{subfigure}[h]{0.24\textwidth}
    \centering
        \includegraphics[width = \textwidth]{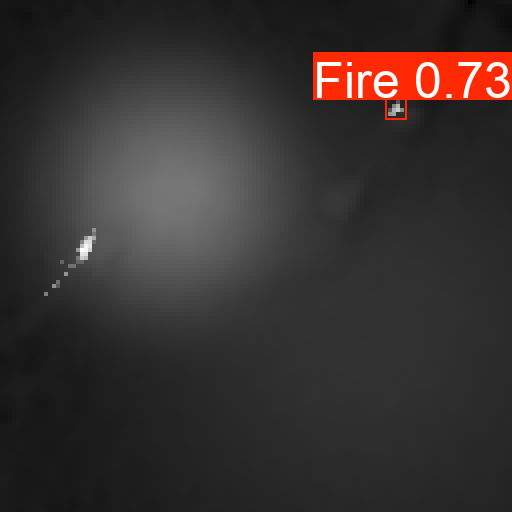}
        \caption{Early Fusion Model interpretation of island image.}
        \label{fig:Interp_Island}
    \end{subfigure}
    \hfill
    \begin{subfigure}[h]{0.24\textwidth}
    \centering
        \includegraphics[width = \textwidth]{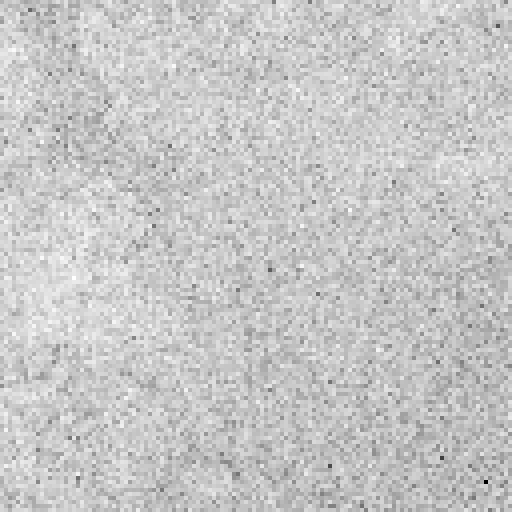}
        \caption{Band $6$ arid region image at night.}
        \label{fig:Band6_China}
    \end{subfigure}
    \hfill
    \begin{subfigure}[h]{0.24\textwidth}
    \centering
        \includegraphics[width = \textwidth]{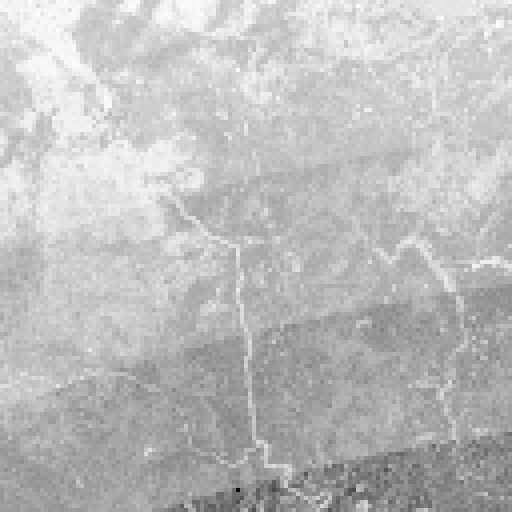}
        \caption{Band $7$ arid region image at night.}
        \label{fig:Band7_China}
    \end{subfigure}
    \hfill
    \begin{subfigure}[h]{0.24\textwidth}
    \centering
        \includegraphics[width = \textwidth]{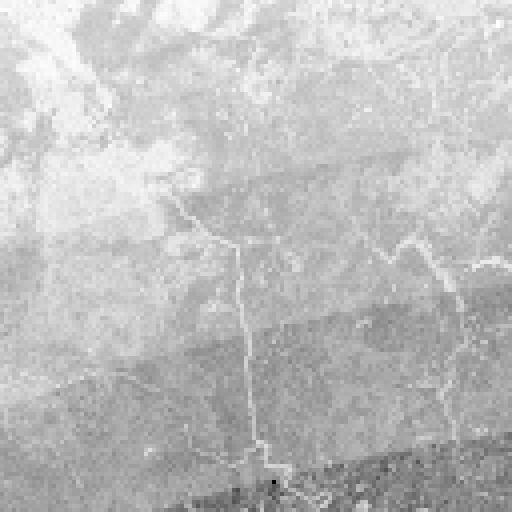}
        \caption{Early fusion arid region image at night.}
        \label{fig:Early_China}
    \end{subfigure}
    \hfill
    \begin{subfigure}[h]{0.24\textwidth}
    \centering
        \includegraphics[width = \textwidth]{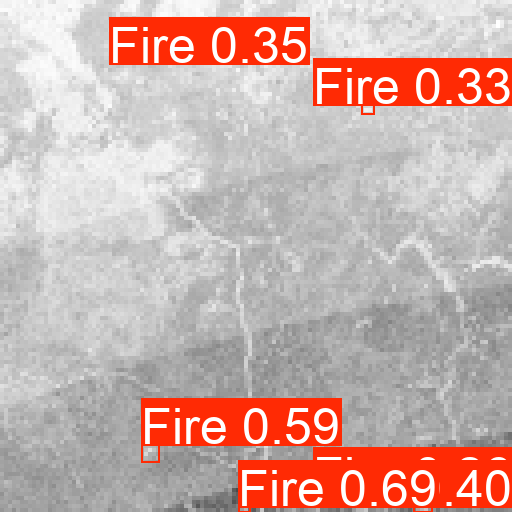}
        \caption{Early Fusion Model interpretation of arid region image.}
        \label{fig:Interp_China}
    \end{subfigure}
    \caption{Examples of false-positive detections.}
    \label{fig:False_Positives}
\end{figure}

Throughout the mission horizon, two constant occurrences have often resulted in false-positive CNN interpretations, ultimately leading to false targets being included in the priority target set. Figure~\ref{fig:False_Positives} shows one example of each consistent case of false-positives, depicting the following imagery from left to right: Band $6$ image, Band $7$ image, Early Fusion image resulting from Algorithm~\ref{Alg:Early_Fusion}, and Early Fusion Model interpretation. Specifically, Figs.~\ref{fig:Band6_Island}--\ref{fig:Interp_Island} show the first case and Figs.~\ref{fig:Band6_China}--\ref{fig:Interp_China} show the second case. The first false-positive occurrence is with small groupings of islands in the ocean, as seen in daytime images. Shown in Figs.~\ref{fig:Band6_Island}--\ref{fig:Early_Island}, the high reflectance of the small islands relative to the dark ocean background produces bright pixels similar to those of wildfires. The corresponding CNN interpretation is visualized in Fig.~\ref{fig:Interp_Island} and depicts a high confidence value of \SI{73}{\%}. This indicates that the spatial contrast seen in maritime geographical areas is alike to the spatial thermal reflectance of wildfires in the simulation, which is a similarity that the current detection module cannot distinguish. The second occurrence of false-positives involves high-reflectance land segments in arid regions at night, as shown in Figs.~\ref{fig:Band6_China}--\ref{fig:Early_China}. These detections are products of the early fusion process, specifically the algorithm's dependence on the Band 7 image. In these geographical areas, the high-reflectance glinting spectrally resembles wildfire pixels. This can be seen in the CNN interpretation in Fig.~\ref{fig:Interp_China}; while the initial interpretations contain several false-positives, the Multi-Pass Confidence Module successfully filters all but the brightest instance on subsequent flyovers. This validates the module's effectiveness in differentiating fusion artifacts from potential false-positive detections from actual active wildfires over time.

\newpage
\section*{Acknowledgments}
This work was supported by the NASA Established Program to Stimulate Competitive Research, Grant\\ \#80NSSC22M0027. The authors thank the anonymous reviewers for their feedback, insight, and thoughtful suggestions, which greatly improved the quality of the manuscript.

\bibliography{References}

\end{document}